\documentclass[conference]{IEEEtran}
\IEEEoverridecommandlockouts
\usepackage{cite}
\usepackage{amsmath,amssymb,amsfonts}
\usepackage{algorithmic}
\usepackage{textcomp}
\usepackage{xcolor}
\usepackage{amssymb}
\usepackage{graphicx}
\usepackage{subfigure}
\usepackage{caption}
\usepackage{graphicx, subfig}
\usepackage{multirow}
\usepackage{array}
\usepackage{amsmath}
\usepackage{mdwmath}
\usepackage{mdwtab}
\usepackage{eqparbox}
\usepackage{url}
\usepackage{xcolor,soul,framed}
\usepackage{pifont}
\usepackage{makecell}
\usepackage{algorithmic}

\begin{document}

\title{CausalRCA: Causal Inference based Precise Fine-grained Root Cause Localization for Microservice Applications\\}

% \author{Ruyue Xin,
%         Peng Chen,
%         Zhiming Zhao% <-this % stops a space
% \IEEEcompsocitemizethanks{\IEEEcompsocthanksitem The authors are with Multiscale Networked Systems (MNS), University of Amsterdam, the Netherlands.\protect
% % note need leading \protect in front of \\ to get a newline within \thanks as
% % \\ is fragile and will error, could use \hfil\break instead.
% %E-mail: see http://www.michaelshell.org/contact.html
% \IEEEcompsocthanksitem Corresponding authors: Peng Chen (chenpeng@mail.xhu.edu.cn) and Zhiming Zhao (z.zhao@uva.nl).
% \IEEEcompsocthanksitem Peng Chen is also with School of Computer and Software Engineering, Xihua University, Chengdu, China.
% }}

% \author{\IEEEauthorblockN{Ruyue Xin\IEEEauthorrefmark{1},
% Peng Chen\IEEEauthorrefmark{2}\IEEEauthorrefmark{3}, 
% Zhiming Zhao\IEEEauthorrefmark{1}\IEEEauthorrefmark{3}}
% \IEEEauthorblockA{
% %Multiscale Networked Systems (MNS) research group, University of Amsterdam \\
% %\\
% \IEEEauthorrefmark{1} Multiscale Networked Systems (MNS) research group, University of Amsterdam \\
% \IEEEauthorrefmark{2} School of Computer and Software Engineering, Xihua University, Chengdu, China \\

% %Email: \IEEEauthorrefmark{1}author.one@add.on.net,
% %\IEEEauthorrefmark{2}author.two@add.on.net,
% %\IEEEauthorrefmark{3}author.three@add.on.net,
% %\IEEEauthorrefmark{4}author.four@add.on.net}
% }}

\author{\IEEEauthorblockN{Ruyue Xin}
\IEEEauthorblockA{\textit{Multiscale Networked Systems (MNS) } \\
\textit{research group, University of Amsterdam}\\
Amsterdam, Netherland \\
}
\and
\IEEEauthorblockN{Peng Chen\IEEEauthorrefmark{1}}
\IEEEauthorblockA{
\textit{School of Computer and Software}\\
\textit{Engineering, Xihua University}\\
Chengdu, China\\
chenpeng@mail.xhu.edu.cn}
\and
\IEEEauthorblockN{Zhiming Zhao\IEEEauthorrefmark{1}}\thanks{\IEEEauthorrefmark{1} Peng Chen and Zhiming Zhao are the corresponding authors}
\IEEEauthorblockA{\textit{Multiscale Networked Systems (MNS) } \\
\textit{research group, University of Amsterdam}\\
Amsterdam, Netherland \\
z.zhao@uva.nl}
}

\maketitle

\begin{abstract}
Effectively localizing root causes of performance anomalies is crucial to enabling the rapid recovery and loss mitigation of microservice applications in the cloud. Depending on the granularity of the causes that can be localized, a service operator may take different actions, e.g., restarting or migrating services if only faulty services can be localized (namely, coarse-grained) or scaling resources if specific indicative metrics on the faulty service can be localized (namely, fine-grained). Prior research mainly focuses on coarse-grained faulty service localization, and there is now a growing interest in fine-grained root cause localization to identify faulty services and metrics. Causal inference (CI) based methods have gained popularity recently for root cause localization, but currently used CI methods have limitations, such as the linear causal relations assumption and strict data distribution requirements. To tackle these challenges, we propose a framework named CausalRCA to implement fine-grained, automated, and real-time root cause localization. The CausalRCA uses a gradient-based causal structure learning method to generate weighted causal graphs and a root cause inference method to localize root cause metrics. We conduct coarse- and fine-grained root cause localization to evaluate the localization performance of CausalRCA. Experimental results show that CausalRCA has significantly outperformed baseline methods in localization accuracy, e.g., the average $AC@3$ of the fine-grained root cause metric localization in the faulty service is 0.719, and the average increase is 10\% compared with baseline methods. In addition, the average $Avg@5$ has improved by 9.43\%.
\end{abstract}


\section{Introduction}
Microservices architecture \cite{balalaie2016microservices} builds cloud applications by decomposing the system functionalities into multiple independently deployable units, making applications more resilient, robust, and adaptable to dynamic cloud environments. The performance of a microservice application is vital to guarantee the quality of user experience (QoE) and service (QoS)\cite{xin2023robust}. However, performance anomalies, such as degraded response time, are inevitable due to the large scale and complex dependencies of services, causing enormous economic loss and user dissatisfaction \cite{li2022actionable}. Furthermore, the performance of applications heavily depends on the underlying resources \cite{gregg2014systems}; for example, high CPU usage results in a congested queue and growing latency \cite{ibidunmoye2015performance}. In order to enable application operations to take actions to resolve performance anomalies effectively, root cause localization to identify faulty services or resources is at the core of software maintenance for online service systems \cite{chen2020towards}. 

A microservice application can be observed by monitoring tools, which help operators to detect performance anomalies\cite{chen2022effectively, Song2023fgcs}. However, performance anomaly detection only notifies operators when an anomaly occurs. To effectively handle performance anomalies, operators need to be informed about where the anomaly occurs (e.g., the faulty service) and what causes the anomaly (e.g., the memory leak). Root causes of a performance anomaly can be localized at different granularity: coarse-grained and fine-grained \cite{chen2014causeinfer}. Coarse-grained means that only faulty services can be identified, and the corresponding action will be the migration or restart of the entire service \cite{wu2021identifying}, which is simple and straightforward but may not recover anomalies and has a higher risk of affecting other services and longer recovery times \cite{wu2020microras}. The developer of Instana Autotrace\footnote{https://www.instana.com/} emphasized the importance of identifying anomaly locations and root causes to avoid delays associated with restarting services, as this may not solve anomalies\cite{InstanaAutoTrace}.

%recent works: mostly focus on coarse-grained; fewer CI methods are used.
At a fine granularity, root cause localization will identify not only the faulty service but also the underlying resources through monitoring metrics of the service \cite{wu2021causal}. Operators can choose accurate actions to mitigate the performance anomaly using fine-grained root cause when pinpointing indicative metrics on the faulty service \cite{chen2014causeinfer, wu2020microrca}. For example, service scale-out has a positive effect and shorter recovery time compared with service restart in the case of underlying resources being insufficient \cite{wu2020microras}. In industry, fine-grained root cause localization attracts much attention. The CCF AIOps Challenge\footnote{http://iops.ai/}, jointly organized by industry and academia, aims to solve problems in real IT operations scenarios based on production systems of industrial companies (e.g., Sougo, eBay, Tencent) \cite{li2022constructing}, providing the performance diagnosis challenge for microservice systems and requires localizing root causes at the metric level in 2020\footnote{https://competition.aiops-challenge.com/home/competition/1484441527290765368}. Instana Autotrace\footnote{https://www.instana.com/},  Google Cloud Operations\footnote{https://cloud.google.com/products/operations} are commercial platforms to identify root causes to help developers and operators fix performance issues \cite{GCOperations}. However, these platforms work with trace data that requires integrating tracking codes into applications and require time and expert technologies to analyze data. Monitoring data, which is different from trace data, can be readily collected and utilized for fine-grained root cause localization in microservice applications, aiding service operators in efficiently and cost-effectively identifying faulty services and pinpointing faulty metrics to resolve performance anomalies \cite{chen2014causeinfer}.

To track the root cause localization problem of microservice applications, some research has developed in recent years. We can classify them into log-based, trace-based, and metric-based according to the data sources \cite{soldani2022anomaly}. Log-based \cite{aggarwal2020localization} and trace-based \cite{kim2013root,weng2018root} research have limitations, such as complex real-time processing and information extraction. On the other hand, metric-based research uses real-time monitoring data, including service latency and system resources, and focuses on localizing faulty services and resource metrics. This kind of research can assist anomaly recovery in taking actions like resource scaling easily without intervention of application source code \cite{soldani2022anomaly}. Nowadays, most metric-based research is coarse-grained \cite{lin2018microscope,guan2018anomaly, wang2018cloudranger, ma2019ms, ma2020self, ma2020automap}, and fine-grained root cause localization is starting to catch the attention of researchers \cite{meng2020localizing, wu2021microdiag}. As for localization methods in metric-based research, causal inference (CI) based methods that can model causal effects between services have been developing recently. For example, CauseInfer \cite{chen2016causeinfer} applies the PC algorithm (named after its authors, Peter and Clark) \cite{spirtes2000causation}, and MicroDiag \cite{wu2021microdiag} uses the linear non-Gaussian acyclic model (LiNGAM) \cite{shimizu2011directLiNGAM} to obtain causal graphs of metrics, which can be seen as anomaly propagation paths. However, currently used CI methods have limitations, such as uncertainty about some causal relations between metrics and strict assumptions about input data and causal relations \cite{wu2021causal}. Therefore, advanced CI methods can be considered for fine-grained root cause localization to discover anomaly propagation paths and improve localization performance.

Fine-grained root cause localization in a microservice application is challenging because 1) services are often heterogeneous and have different characteristics, which may result in diverse anomaly symptoms for the same issue; 2) the complex dependency between microservices makes it difficult to model the anomaly propagation resulting from faulty services; 3) a large number of anomalous metrics introduced in a system makes it hard to find out the root one for a performance anomaly. To address these challenges, we formulate our main research question: \textit{how to pinpoint the root cause of performance anomalies at a fine granularity based on monitoring data?} Three sub-questions are proposed: \textit{1) how to model anomaly propagation between monitoring metrics using CI methods? 2) How to precisely determine the root cause based on the propagation model? 3) How to evaluate the performance of root cause localization result?}

To answer the research question, we propose a CI-based fine-grained root cause localization framework named CausalRCA for microservice applications in this paper. The framework works when an anomaly is detected. Based on real-time monitoring data, CausalRCA will perform automatic root cause localization, including modeling anomaly propagation paths as a causal graph and ranking metrics to localize the root cause by traversing along the graph. Finally, CausalRCA outputs predicted root causes, which can be used by operators to determine strategies and recovery actions to solve the anomaly. In this paper, we evaluate the localization performance of CausalRCA on the sock-shop\footnote{https://github.com/microservices-demo/microservices-demo\label{sock}} microservice benchmark. When a performance anomaly in sock-shop, such as the high response time of user requests, is detected, we can input monitoring metrics, including service latency and resource metrics of each service, to CausalRCA. After processing, the faulty service and root cause metric, for example, the memory usage metric in the order service, will be identified. Our experimental results show that CausalRCA has improved localization accuracy. For example, the average improvement of $AC@5$ for the fine-grained root cause metric localization in the faulty service is 9.43\% compared with baseline methods.  

Our contributions can be summarized below: 
%Research question: how to localize the root cause accurately? 
%Therefore, we consider the DAG-GNN method. 
\begin{itemize}[noitemsep, nolistsep]
\item We propose an automated, fine-grained root cause localization framework named CausalRCA, which analyzes monitoring data and localizes faulty services and system resources in real-time.  
\item We provide a gradient-based causal structure learning method in CausalRCA, which can automatically capture linear and non-linear causal relations between monitoring metrics. 
\item We conduct coarse- and fine-grained experiments to evaluate the localization performance of CausalRCA and demonstrate that the proposed framework has the best localization accuracy compared with baseline methods. For example, the average $AC@3$ is 0.719, which is a 10\% improvement compared with baseline methods.
\end{itemize}

The rest of the paper is organized as follows. In Section \ref{relw}, we review existing root cause localization research and CI methods. In Section \ref{framw}, we propose a framework for root cause localization and a detailed introduction of each method. In Section \ref{exp}, we design experiments from coarse-grained to fine-grained to evaluate the localization performance of our framework. Finally, discussion and conclusion are provided in Section \ref{dis} and Section \ref{con}.

\section{Related works}
\label{relw}
In recent years, research has developed for root cause localization in distributed system \cite{gholami2021comparative}, clouds \cite{weng2018root, wu2020microrca}. Based on data sources, we can categorize these researches into three groups: log-based, trace-based, and metric-based \cite{soldani2022anomaly}. Log-based research \cite{aggarwal2020localization} mainly localizes root causes based on text logs parsing, which is hard to work in real time. Trace-based research \cite{kim2013root,weng2018root} gathers information through the complete tracing of the execution paths and then identifies root causes along those paths. However, trace data only focuses on service level, and it is time-consuming for developers to understand source code well enough to extract trace information. In contrast, metrics-based research uses monitoring data collected from applications and underlying infrastructures to construct causal graphs and infer root causes. Metric-based research can achieve automated, real-time root cause localization based on multi-dimensional information. Therefore, this section will mainly review metrics-based research and CI methods. 

\subsection{Metric-based root cause localization research}

%Trace-related: \cite{yu2021microrank}
Based on monitoring data, some researchers identify root causes of performance anomalies with statistical analysis, e.g., identifying anomalous monitoring metrics in parallel with detected anomalies. Want et al.\cite{wang2020root} conduct correlation analysis based on mutual information to determine the root-cause metric for the anomalies they detect. However, given that correlation does not ensure causation \cite{calude2017deluge}, statistical analysis cannot pinpoint root causes. In addition, some researchers have developed topology graph-based analysis, which reconstructs the topology graph of a running application. For example, Wu et al.\cite{wu2020microrca} generate a topology graph based on deployment information and extract a weighted anomalous subgraph by parsing resource-level monitoring data. Brandón et al.\cite{brandon2020graph} make snapshots of abnormal states of the application as graphs and then identify the root cause of a new anomaly by graph matching. This kind of research uses a reconstructed application topology graph to determine root causes, which can only be used for coarse-grained root cause localization. 

To identify related papers for our research on CI-based root cause localization, we conduct a thorough search using Google Scholar. We use specific keywords such as "microservice", "causal inference", and "root cause localization" to narrow down the results, and limit the search to recent papers published between 2012 and 2022. We sort the papers based on relevancy and identified the two most related papers \cite{meng2020localizing, wu2021microdiag}. We carefully examine these papers and their benchmark methods and discover CauseInfer \cite{chen2014causeinfer}, which is a well-respected and influential research work. The CauseInfer focused on fine-grained real reasons causing performance problems, introducing a low-cost, black box cause inference system to build a causality graph and infer the causes from the graph. We use the snowballing technique \cite{wohlin2014guidelines} to explore CauseInfer's citation papers to identify valuable papers and establish a comprehensive understanding of related research in CI-based root cause localization. We classify these research works into coarse- and fine-grained categories and summarize them chronologically, as shown in Table \ref{tab:cal-graph}.

Existing CI-based root cause localization research works by constructing a causal graph based on monitoring data, i.e., including causal structure learning and root cause inference \cite{soldani2022anomaly}. Coarse-grained root cause localization usually builds a causal graph based on service level objective (SLO) metrics, such as service latency, and focuses on determining faulty services. For example, Microscope \cite{lin2018microscope,guan2018anomaly} collect information on service interactions and monitoring service latency and then processes them based on PC and breadth-first search (BFS) algorithms to determine possible faulty service of detected anomalies. In addition, CloudRanger \cite{wang2018cloudranger}, MS-Rank \cite{ma2019ms,ma2020self}, and AutoMAP \cite{ma2020automap} all exploit PC and random walk algorithms to build causal graphs and infer root causes. MS-Rank and AutoMAP use metrics not only service latency but also throughput, power, and resource consumption, whereas AutoMAP develops novel operations to refine the causal graph. Various coarse-grained root cause localization studies have been conducted, but they cannot assist operators in resolving application anomalies with accurate actions. 

To address the drawback of coarse-grained root cause localization, some researchers focus on fine-grained root cause localization. Chen et al. first proposed CauseInfer \cite{chen2014causeinfer,chen2016causeinfer}, which infers the faulty service and root cause metric by constructing a causality graph of monitoring metrics in each service with the PC algorithm and traversing the metric causality graph with a depth-first search (DFS) method. After several years, Meng et al. provided MicroCause \cite{meng2020localizing}, which mainly focuses on the root cause metric localization in a faulty service. It provides a PC-based causal graph building method for time-series data and infers root causes with the random walk method. Afterward, Wu et al. proposed MicroDiag \cite{wu2021microdiag}, which focuses on fine-grained root cause localization and applies a DirectLiNGAM to build causal graphs and PageRank to infer root causes. Fine-grained root cause localization focuses mainly on SLO metrics and monitoring resource metrics of services and determining the root cause resource metric to help operators take actions like scaling resources \cite{tuli2022pregan}. 

\begin{table}[htb!]
\centering
\caption{\label{tab:cal-graph} Classification of metric-based root cause localization research} 
\scalebox{0.78}{
\begin{tabular}{|p{1.5cm}|p{0.6cm}|p{2.0cm}|p{1.2cm}|p{1.0cm}|p{1.0cm}|p{1.0cm}|}
\hline
\makecell[c]{\textbf{Reference}} & 
\makecell[c]{\textbf{Year}} & 
\makecell[c]{\textbf{Causal structure}\\ \textbf{learning}} &
\makecell[c]{\textbf{Root cause}\\ \textbf{inference}} &
\makecell[c]{\textbf{Input}} &
\makecell[c]{\textbf{Root} \\ \textbf{cause}} &
\makecell[c]{\textbf{Granu-} \\ \textbf{larity}}\\ \hline
Micorscope\cite{lin2018microscope,guan2018anomaly} & 2018 & Parallelized PC & BFS & Service latency & Faulty service & Coarse-grained \\ \hline
CloudRanger\cite{wang2018cloudranger} & 2018 & PC & Random walk & Service latenct & Faulty service & Coarse-grained \\ \hline
MS-Rank\cite{ma2019ms,ma2020self} & 2019 & PC & Random walk  & Multi metrics & Faulty service & Coarse-grained \\ \hline
AutoMAP\cite{ma2020automap} & 2020 & PC & Random walk & Multi metrics & Faulty service & Coarse-grained \\ \hline
%A RCA method\cite{qiu2020causality} & PC+KG, BFS & monitored KPIs & ranked services metrics  \\ \hline
CauseInfer\cite{chen2014causeinfer,chen2016causeinfer} & 2014, 2016 & PC & BFS & Service latency & Faulty service & Fine-grained \\ \hline
MicroCause\cite{meng2020localizing} & 2021 & PCTS & Random walk & Resource metrics & Root cause metric & Fine-grained \\ \hline
MicroDiag\cite{wu2021microdiag} & 2021 & DirectLiNGAM  & PageRank & Service and resource metrics & Root cause metric & Fine-grained \\ \hline
\end{tabular}}
\end{table}

In Table \ref{tab:cal-graph}, we can see that much research is focusing on coarse-grained root cause localization. However, fine-grained root cause localization was proposed early and has begun to attract the attention of more researchers in recent years. As for CI methods, we can see that causal structure learning methods like PC and LiNGAM are applied. At the same time, root cause inference methods, such as BFS and random walk, are popular. Based on these works, we consider precise root cause localization is more helpful for microservice application recovery from performance anomalies. At the same time, the localization accuracy of existing works can be improved; for example, the success rate of accurately identifying the root cause may be under 20\% \cite{wang2018cloudranger, lin2018microscope}. Therefore, we are motivated to explore fine-grained root cause localization with advanced CI methods. 

\subsection{Causal inference methods}
Causal inference methods, especially causal structure learning, have been researched for several years, and they play a vital role in many areas, such as genetics \cite{peters2017elements} and biology \cite{sachs2005causal}. The causal structure learning problem can be formulated to learn a directed acyclic graph (DAG) from observational data. Methods can be classified into constrained-based, score-based, function-based, and gradient-based. Constrained-based methods, such as PC and FCI \cite{spirtes2000causation}, use conditional independence tests to learn the skeleton of the casual graph and then orient the edges based on pre-defined orientation rules. Score-based methods, like GES \cite{chickering2002optimal}, assign scores to different causal graphs based on a pre-defined score function and then search over the space of DAGs to find the optimal one. Finally, function-based methods, like LiNGAM \cite{shimizu2006linear, shimizu2011directLiNGAM}, construct a linear structural equation model (SEM) based on linear and non-Gaussian assumptions, and solve it to get the DAG. These traditional methods contribute much to causal structure learning but have limitations. PC usually has ambiguous causal relations in causal graphs. GES takes a long time to match graphs, which makes it inappropriate to be applied to large-scale data. LiNGAM has strict linear and non-Gaussian assumptions, which makes it impractical. 

With the development of deep neural networks, gradient-based methods are developed. Zheng et al.\cite{zheng2018dags} propose an equality constraint to the linear SEM, which enables a suite of continuous optimization techniques such as gradient descent. After that, Yu et al.\cite{yu2019dag} provide a deep generative model and apply a variant of the structural constraint to learn the DAG. Gradient-based methods have no limitation of input data, can deal with linear and non-linear causal relations in data, and can automatically generate a weighted DAG. Gradient-based methods have been applied to medical \cite{upadhyaya2021scalable} and biology \cite{demir2022eeg}. However, to the best of our knowledge, no research has applied gradient-based methods to root cause localization of microservice applications. 

Based on DAGs generated by causal structure learning methods, researchers apply graph methods, like BFS \cite{beamer2012direction}, random walk \cite{spitzer2001principles}, and PageRank \cite{ridings2002pagerank}, for root cause inference. The BFS is to traverse the graph and determine the abnormal node without descendants or with no abnormal descendants as a root cause. A random walk is walking through paths and choosing neighbors randomly in a graph. It determines the node most visited as the root cause. PageRank improves the random walk by adding the possibility of jumping to a random node, which will be used in this paper. 

In conclusion, metric-based research can achieve automated and real-time root cause localization compared with log- and trace-based research. However, most existing research is about coarse-grained faulty service localization, while fine-grained root cause metric localization can be more helpful for rapid recovery and loss mitigation. CI-based methods are popular, but currently used methods have their limitations. Therefore, this paper will mainly focus on fine-grained root cause localization and explore gradient-based methods to build causal graphs.

\section{Root cause localization framework}
\label{framw}

In this section, we propose a root cause localization framework named CausalRCA, including causal structure learning and root cause inference, and we will introduce detailed methods in the framework. All codes and data can be found in our Github repository CausalRCA\footnote{https://github.com/AXinx/CausalRCA\_code.git}. 

\begin{figure*}[htbp]
\centering
\includegraphics[width=0.8\textwidth]{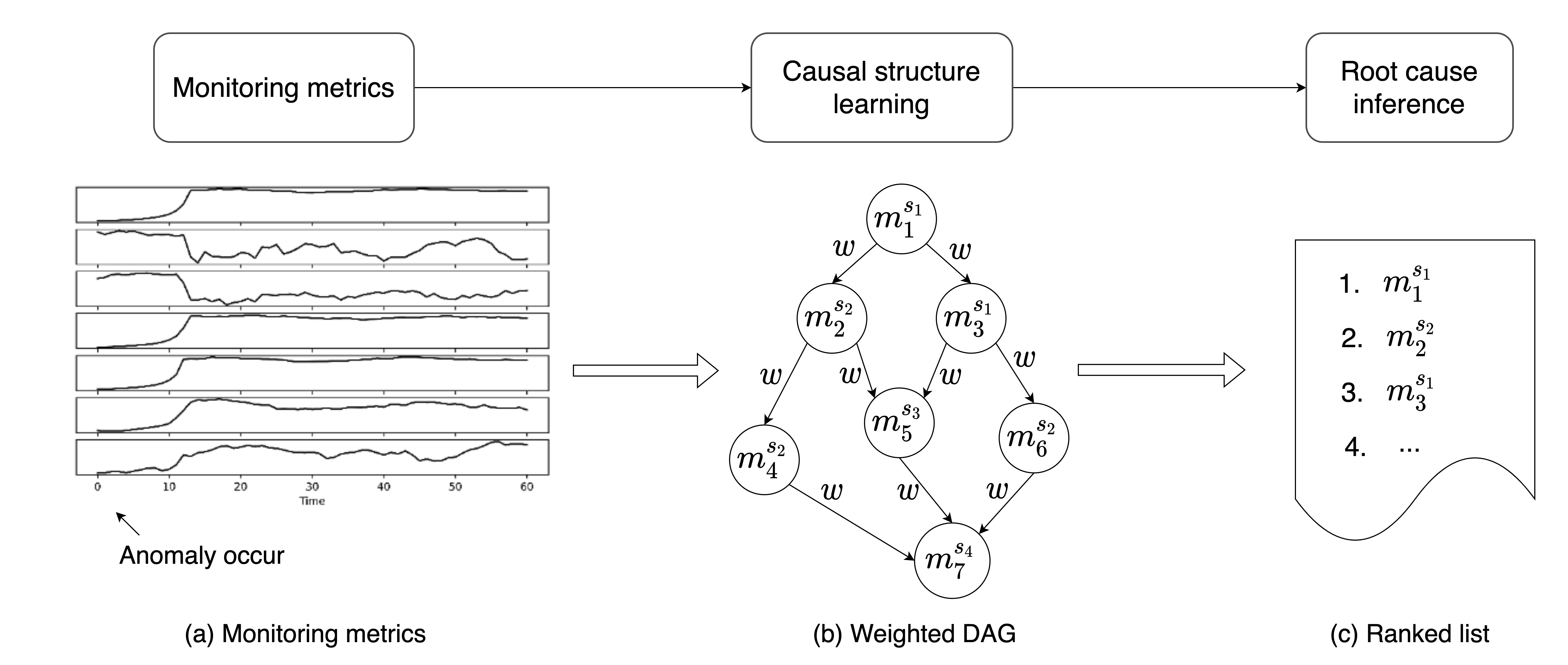}
\caption{CausalRCA: details of the root cause localization framework}
\label{fig:fram}
\end{figure*}

\subsection{Framework overview}

The CausalRCA can automatically build anomaly propagation paths and localize root causes in real-time based on observable metrics. The CausalRCA framework consists of three components: monitoring metrics, causal structure learning, and root cause inference, as shown in Figure \ref{fig:fram}.

The CausalRCA works when an anomaly occurs, such as the high latency of user requests, and it then automatically build anomaly propagation paths and localize root causes in real-time based on observable metrics. We first collect monitoring data, including service-level data, that is, service latency, and resource-level data, such as container CPU/memory usage. We use $m_i^{s_j}$ to represent a monitoring metric in the service $s_j$, and all monitoring data is time-series data as shown in Figure \ref{fig:fram}(a). Based on monitoring data, we then start the causal structure learning. The causal structure learning will automatically build a causal graph of metrics, which can be seen as anomaly propagation paths. We develop the causal structure learning with a gradient-based CI method, which can output a weighted DAG to represent causal relations between metrics as shown in Figure \ref{fig:fram}(b). With the DAG, we start root cause inference to localize root causes. We apply PageRank to the weighted DAG and output a ranked list of all metrics, as shown in Figure \ref{fig:fram}(c). Depending on the input data, the CausalRCA can be used for coarse- or fine-grained root cause localization. Coarse-grained works when input service latency, and CausalRCA will output the faulty service. Fine-grained works when input resource metrics, and CausalRCA will output the root cause metric. We evaluate the localization performance of CausalRCA in experiments in Section \ref{exp}.

\subsection{Causal structure learning}
The causal structure learning component aims to build a causal graph of monitoring metrics. The causal graphs can be seen as anomaly propagation paths between metrics. We can use a DAG to represent the causal graph, in which each node represents a metric, and each edge represents a cause-effect relationship. Based on related work, we know that traditional causal structure learning methods have strict limitations on input data and relations. Therefore, we implement the causal structure learning in CausalRCA with a gradient-based method, DAG-GNN \cite{yu2019dag}. DAG-GNN provides a deep generative model, which is a variational autoencoder (VAE) parameterized by a novel graph neural network (GNN) \cite{bruna2013spectral}, and applies a variant of the structural constraint to learn DAGs. Unlike other causal structure learning methods, the gradient-based method has no limitation of input data, can extract linear or non-linear causal relations between metrics, and automatically outputs a weighted DAG. 

We use $X\in\mathbb{R}^{m\times{n}}$ ($m$ is metrics, $n$ is samples of each metric) to represent input data. To get a DAG from $X$, Zheng et al.\cite{zheng2018dags} adopt a linear SEM as a data generation model, which is $X = A^TX+Z$ ($A\in\mathbb{R}^{m\times{m}}$ is the weighted adjacency matrix. $Z\in\mathbb{R}^{m\times{n}}$ is the noise matrix). To ensure the acyclicity of the DAG, a constraint of A is proposed as:
\begin{equation}
    h(A) = tr\left[(I+\alpha{A \circ A})^m\right] - m  = 0
\end{equation}

\begin{figure}[htbp]
\centering
\includegraphics[width=0.5\textwidth]{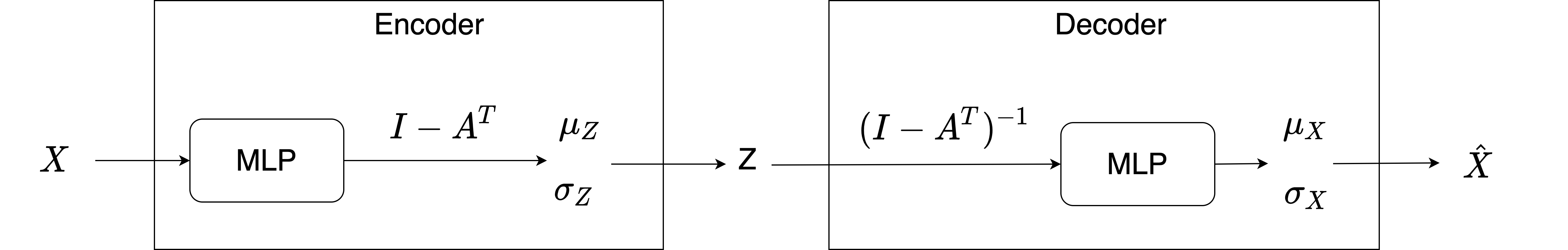}
\caption{Architecture of the causal structure learning method}
\label{fig:gnn}
\end{figure}

Based on the linear SEM, we can get $X = (I-A^T)^{-1}Z$, which can be written as $X = f_A(Z)$. This equation is a general form recognized as an abstraction of parameterized GNNs \cite{kipf2016semi}. We can also see that X is generated from a latent representation Z by defining a probabilistic graphical model. The generative model can be developed based on a VAE, and Z follows a standard Gaussian distribution \cite{kingma2013auto} as shown in Figure \ref{fig:gnn}. 

With latent representation Z, we can define the decoder to reconstruct X as: 
\begin{equation}
    X = f_2((I-A^T)^{-1}f_1(Z))
\end{equation}

Then, the corresponding encoder can be defined as:

\begin{equation}
    Z = f_4((I-A^T)f_3(X))
\end{equation}

Combining with deep neural networks, we use multilayer perceptron (MLP) to simulate $f_1$, $f_2$, $f_3$, and $f_4$, which all are parameterized functions. Based on VAE, the output of encoder and decoder are data distributions, so we get $Z$ by sampling from $\mu_Z$ and $\sigma_Z$, and $\hat{X}$ by sampling from $\mu_X$ and $\sigma_X$.

For a VAE model, with a variational posterior $q(Z|X)$ to approximate the actual posterior $p(Z|X)$, evidence lower bound (ELBO) can be represented as: 
\begin{equation}
\begin{aligned}
L_{ELBO} &= E_{Z\sim{q}}\left[\log{p(X|Z)}\right] - KL(q(Z|X), p(Z)) \\
         &= E_{Z\sim{q}}\left(-\frac{1}{2c}\|X - \hat{X}\|\right) - KL(q(Z|X), p(Z))
\end{aligned}
\end{equation}

Thus, the learning problem can be defined as:
\begin{equation}
\begin{aligned}
\mathop{\min}_{A,\theta} f(A,\theta) &= -L_{ELBO} \\
s.t.\ h(A) &= 0 \\
\end{aligned}
\end{equation}

where $\theta$ is all the parameters of the VAE. For a nonlinear equality-constrained problem, we can use augmented Lagrangian method \cite{bertsekas1997nonlinear} to solve it. 

\begin{equation}
L_c(A,\theta,\lambda) = f(A,\theta)+\lambda h(A)+\frac{c}{2}|h(A)|^2
\end{equation}

where $\lambda$ is the Lagrange multiplier and $c>0$ is the penalty parameter. The following update rules are defined: 
\begin{equation}
\begin{aligned}
A^k, \lambda^k &= \mathop{\arg\min}_{A,\theta}L_{c^k}(A,\theta,\lambda^k) \\
\lambda^{k+1} &= \lambda^k + c^kh(A^k) \\
c^{k+1} &= \left\{
\begin{aligned}
\eta c^k,&\ if\ |h(A^k)|>\gamma|h(A^{k-1})| \\
c^k,&\ otherwise
\end{aligned}
\right.
\end{aligned}
\end{equation}

In the augmented Lagrangian, the penalty parameter $c$ is typically updated using an exponentially increasing function of the iteration number, and the Lagrange multiplier $\lambda$ is correspondingly updated to converge to the optimal condition. The update rule for the penalty parameter $c$ is important to balance the trade-off between feasibility and optimality in the optimization problem. The rule states that if the constraint violation at the next iteration is larger than the current violation, the value of $c$ should be increased. Conversely, if the constraint violation at the next iteration is smaller than the current violation, the value of $c$ should be kept the same. To achieve faster convergence and find optimal solutions, the update rule depends on two tuning parameters, $\eta$ and $\gamma$. Usually, we set $\eta >1$ to induce fast convergence and $\gamma <1$ to limit the convergence speed \cite{yu2019dag}. If $\gamma$ is set too high, the convergence will be slow, while if $\eta$ is set too high, the convergence will be fast, but the results may oscillate. Parameter analysis is provided in our experiments in Section \ref{exp_res}.

During training, parameters $A$ and $\theta$ will be updated every epoch. After training, we can get the $A$, which is the adjacency matrix of a DAG. For root cause localization in microservice applications, we define $X=[m_1^{s_1}, m_2^{s_1},...,m_i^{s_j},...]$. With this causal structure learning method, we can get a weighted DAG ($G$) which represent causal relations between metrics as shown in Figure \ref{fig:fram}(b). Each node in $G$ represent a metric, for example, $m_1^{s_1}$ means a metric in service $s_1$. The edge from $m_1^{s_1}$ to $m_2^{s_1}$ indicates that a change in $m_1^{s_1}$ will result in a change in $m_2^{s_1}$ with the weight $w$. Weight $w$ represents the degree of the impact. If $w$ is large, it indicates that a small change in $m_1^{s_1}$ will result in a large change in $m_2^{s_1}$. Furthermore, $w$ can be either negative or positive, implying that an increase in $m_1^{s_1}$ may result in an increase or decrease in $m_2^{s_1}$. Based on the weighted DAG, we then use a root cause inference method to pinpoint the root cause metric. 

\subsection{Root cause inference}

For the weighted DAG ($G$), we can rank metrics with the PageRank algorithm. PageRank works according to the number of incoming edges and the probability of anomalies spreading through the graph. We define $P_{ij}$ as the transition probability of node $i$ to $j$: 

\begin{equation}
P_{ij} = \left\{
\begin{aligned}
\frac{w_{ij}}{\sum_j{w_{ij}}},&\ if\ w_{ij} \neq 0 \\
0,&\ otherwise
\end{aligned}
\right.
\end{equation}

Here, $w_{ij}$ is the weight between node $i$ and $j$. We define $P$ as the transition probability matrix. Then, we can get the PageRank vector $v$ as proposed by \cite{page1999pagerank} as:

\begin{equation}
v = \alpha{P}v + \frac{1-\alpha}{n}
\end{equation}

Here, $n$ is the number of nodes, $\alpha\in(0,1)$ is the teleportation probability, and it means that the random walk will continue with probability $\alpha$ and jump to a random node with probability $1-\alpha$. We use the default setting $\alpha=0.85$ \cite{becchetti2006distribution}. To get results of the root cause inference method better, we first reverse edges in $G$ and use the absolute value of all weights. After running the root cause inference method, we rank the PageRank scores of all nodes and get the ranked list as shown in Figure \ref{fig:fram}(c). The higher the ranking on the list, the more likely the root cause is. 

\section{Experiments and results}
\label{exp}

To evaluate the root cause localization framework CausalRCA, we conduct experiments on both coarse-grained and fine-grained. As for coarse-grained root cause localization, we design experiments to identify faulty services. As for fine-grained root cause localization, we first localize root cause metrics in the faulty service. In addition, taking into account the lack of understanding of services and underlying infrastructures of an application, we provide another fine-grained experiment to localize the root cause metric with all monitoring metrics in all services. In this section, we will introduce experimental settings and experimental results. 

\subsection{Experimental settings}

\subsubsection{Testbed} 
To evaluate our framework, sock-shop\textsuperscript{\ref{sock}}, which simulates an e-commerce website that sells socks, is deployed. It is widely used as a microservice benchmark designed to aid demonstration and test microservices and cloudnative technologies \cite{lin2018microscope, wu2020microrca, wu2021causal}. Sock-shop consists of 13 services, which are implemented in heterogeneous technologies and communicate via REST over HTTP. Except for communication services, it contains 7 functional services, which are, \textit{frontend} serves as the entry of user requests; \textit{catalogue} provides product catalogue and information; \textit{carts} holds shopping carts; \textit{user} stores user accounts, including paymenet cards and addresses; \textit{orders} place orders of login users from carts, and it consumes memory a lot; finally, \textit{payment} and \textit{shipping} services are provided for orders, which require network for processing transactions.

\begin{figure}[htbp]
\centering
\includegraphics[width=0.45\textwidth]{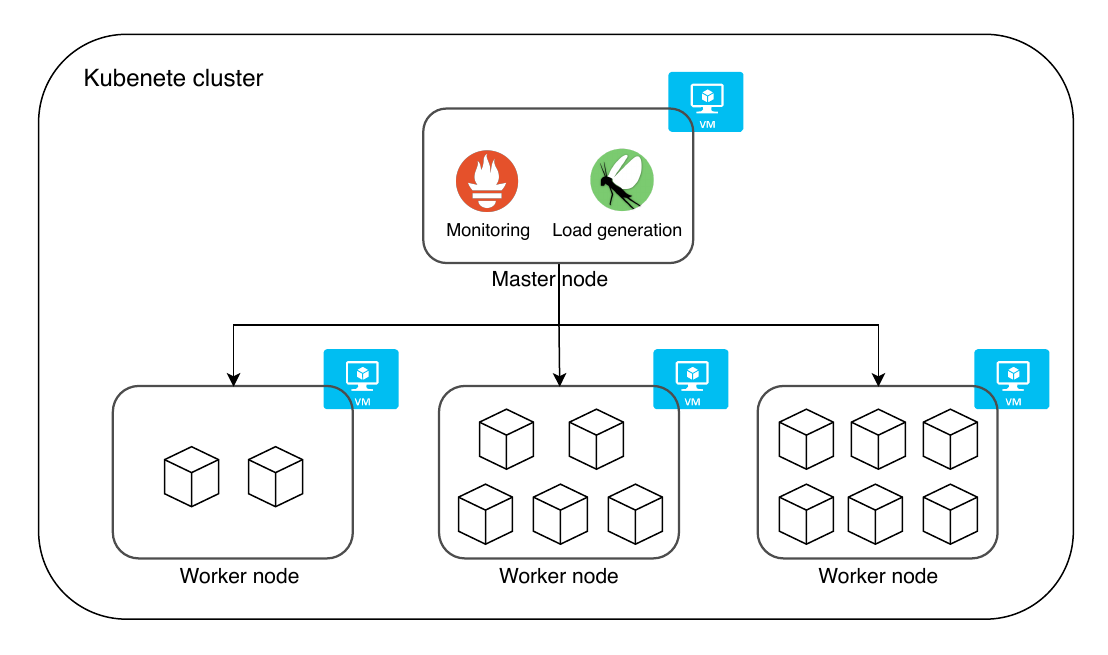}
\caption{The microservice application sock-shop deployed on VMs with Kubernetes}
\label{fig:deploy}
\end{figure}

We deploy the sock-shop with Kubernetes on several VMs in the cloud, as shown in Figure \ref{fig:deploy}.
In the Kubernetes cluster, we have one master node and three worker nodes. Their configurations are Ubuntu 18.04, 4vCPU, 16G RAM Memory, and 80G Disk. On the master node, we deploy open-source monitoring and visualization tools, Prometheus\footnote{https://github.com/prometheus/prometheus} and Grafana\footnote{https://github.com/grafana/grafana}, repectively. Prometheus and Grafana are widely used for monitoring in microservice applications \cite{waseem2021design, promGraf}. Prometheus can keep monitoring the whole system and collecting both service-level and resource-level data \cite{wu2020microrca}. In addition, a load generation tool, Locust\footnote{https://locust.io/}, is deployed on the master node to simulate workloads for the microservice application. On worker nodes, we deploy 13 services of the sock-shop application, and they are allocated to different VMs automatically by Kubernetes.

\subsubsection{Anomaly injection} 

Microservice applications are deployed and distributed in clouds, and their performance is highly dependent on the resources of the underlying infrastructures. There are several common and widespread real performance anomalies in distributed systems  \cite{mariani2018localizing}. Anomalous CPU consumption in VMs due to infinite loops, busy waits, or deadlocks of competing actions in applications can cause a slowdown of user requests  \cite{sauvanaud2016anomaly}. Memory leak, one of the most prominent software bugs that severely threaten the availability and security of systems  \cite{kang2010peerwatch}, happen when allocated chunks of memory are not freed after their use. Accumulations of unfreed memory may exhaust the system resource and lead to memory shortage and system failures. In addition, network resources are vulnerable to being attacked because of the frequent communication between servers and clients. Network latency anomalies usually originate from queuing or processing delays of packets on gateways \cite{sauvanaud2016anomaly}. The three anomalies are commen and frequent in micorservice application \cite{wu2020microrca, lin2018microscope}, which will be used to evaluate our framework. 

Our method can be applied to any anomaly that manifests as increased microservice response time. In this evaluation, we inject the three common anomalies: CPU hog, memory leak, and network delay. We inject CPU hog by consuming CPU resources of each service. For memory leak, we allocate memory continuously for each service. For network delay, we enable traffic control to delay the network packets. We implement anomaly injection with the tool Pumba\footnote{https://github.com/alexei-led/pumba}, which can emulate network failures and stress-testing resources for Docker containers. Based on anomaly detection research \cite{jayathilaka2018detecting, huang2022semi}, anomalies usually last several minutes, so each anomaly of each service we injected lasts 5 minutes, and the application will have 10 minutes to cold down before another injection. 

\subsubsection{Data collection}
We deploy Prometheus to monitor the microservice application and collect monitoring data in real-time. Prometheus is configured to collect data every 5 seconds. We collect both service-level and resource-level data. At the service level, we collect the latency of each service. At the resource level, we collect container resource-related metrics, including CPU usage, memory usage, disk read and write, and network receive and transmit bytes, as shown in Table \ref{tab:metrics}.

\begin{table}[htb!]
\centering
\scalebox{0.9}{
\begin{tabular}{|c|p{3.5cm}|}
\hline
\multicolumn{1}{|c|}{\textbf{Metrics type}} & \multicolumn{1}{c|}{\textbf{Metrics}}\\ \hline
Service-level  & Service latency \\ \hline
\multirow{6}{*}{Resource-level}& CPU usage \\ \cline{2}
~ & Memory usage \\ \cline{2}
~ & Disk read \\ \cline{2}
~ & Disk write \\ \cline{2}
~ & Network receive \\ \cline{2}
~ & Network transmit \\ \hline
\end{tabular}}
\caption{\label{tab:metrics} Collected monitoring metrics}
\end{table}

\subsubsection{Baseline methods}
Related work in Section \ref{relw} shows that CI-based root cause localization uses different causal structure learning methods. Our CausalRCA is developed based on a gradient-based causal structure learning method. Therefore, to evaluate the localization performance of our CausalRCA, we design baseline methods by combining different causal structure learning methods with PageRank. We chose the constraint-based method PC, the score-based method GES, and the function-based method LiNGAM. 

For these baseline structure learning methods, we use their default parameter settings in causal-learn\footnote{https://github.com/cmu-phil/causal-learn}. In CausalRCA, we use 2-layers MLP in the encoder and decoder, respectively. We set the learning rate as $1e-3$, and training epochs as 1000. In addition, we train the model with the Adam optimizer. We use $\eta = 10$ and $\gamma = 0.25$ as default in our experiments, which is proven to work well in \cite{yu2019dag}, and perform parameter analysis with $\eta = 100, 1000$ and $\gamma = 0.5, 0.75$. We run CausalRCA 10 times and take the average as the result of each experiment. 

\subsubsection{Evaluation metrics}
To evaluate localization accuracy, we use two performance metrics: $AC@k$ and $Avg@k$, which are the most commonly used metrics to evaluate rank results\cite{wu2020microrca}. $AC@k$ represents the probability that the top $k$ results given by a method include the real root cause. When the $k$ is small, a higher $AC@k$ indicates that the method is more accurate in identifying the root cause. For each service and each anomaly type, $AC@k$ is calculated as follows:
\begin{equation}
 AC@k = \frac{\sum_{i<k}R{[i]}\in{V_{rc}}}{min(k,|V_{rc}|)}
\end{equation}
where $R[i]$ is the result after ranking all metrics. $V_{rc}$ is the root cause set. $Avg@k$ evaluates the overall performance of a method by computing the average $AC@k$, which is defined as:
\begin{equation}
Avg@k = \frac{1}{k}\sum_{1\leq{j}\leq{k}}AC@j
\end{equation}
We use AC@1, AC@3, and Avg@5 in our experiments. AC@1 evaluates if the top localized root cause is the real one, and it is the most restrictive and accurate metric. AC@3 is used to determine if the top three localized results have the real root cause. This metric is less accurate than $AC@1$, but it can still help operators quickly reduce root-cause candidates and localize the real ones. Finally, Avg@5 represents the average localization ability. The three metrics are commonly used in the root cause localization task, and they can fairly evaluate localization performance \cite{wu2020microrca, meng2020localizing}.

\subsubsection{Statistical testing}

To assess the statistical significance of different RCA methods, we use the one-way analysis of variance (ANOVA) to test the difference between all RCA methods and the t-test to check the pairwise differences \cite{demvsar2006statistical}. We use $Avg@5$ as the performance score of each RCA method. ANOVA is a hypothesis-testing framework for determining whether the between-group variation is significant. The F-statistic, calculated as the ratio of the between-group variation to the within-group variation, is used in ANOVA. The p-value associated with the F-statistic indicates the probability of obtaining an F-statistic as extreme as the observed one, assuming the null hypothesis is true. If the p-value is less than the significance level (usually 0.05), we reject the null hypothesis and conclude that there is a statistically significant difference among the RCA methods.

If the ANOVA test indicates a significant performance difference among these methods, we then use a t-test to determine the differences between each pair of RCA methods. The basic idea behind the t-test is to calculate a test statistic called the t-value, which measures the difference between the average performance scores of two methods relative to the variability within each method. Once the t-value has been calculated, it is compared to a critical value from a t-distribution. If the t-value is less than the critical value, we fail to reject the null hypothesis and conclude that there is insufficient evidence to suggest a performance difference between the two methods.

\subsection{Experimental results}
\label{exp_res}

We provide the results of three experiments as below:
\begin{itemize}[noitemsep, nolistsep]
\item Coarse-grained faulty service localization based on service latency of all services. 
\item Fine-grained root cause metric localization in the faulty service based on system-level metrics in the faulty service.
\item Fine-grained root cause metric localization with all monitoring metrics in all services
\end{itemize}
We compare the localization performance of CausalRCA with baseline methods and explain the results.

\subsubsection{Coarse-grained faulty service localization}

\begin{figure}[htbp]
\centering
\begin{minipage}[t]{0.4\textwidth}
\centering
\includegraphics[width=6cm]{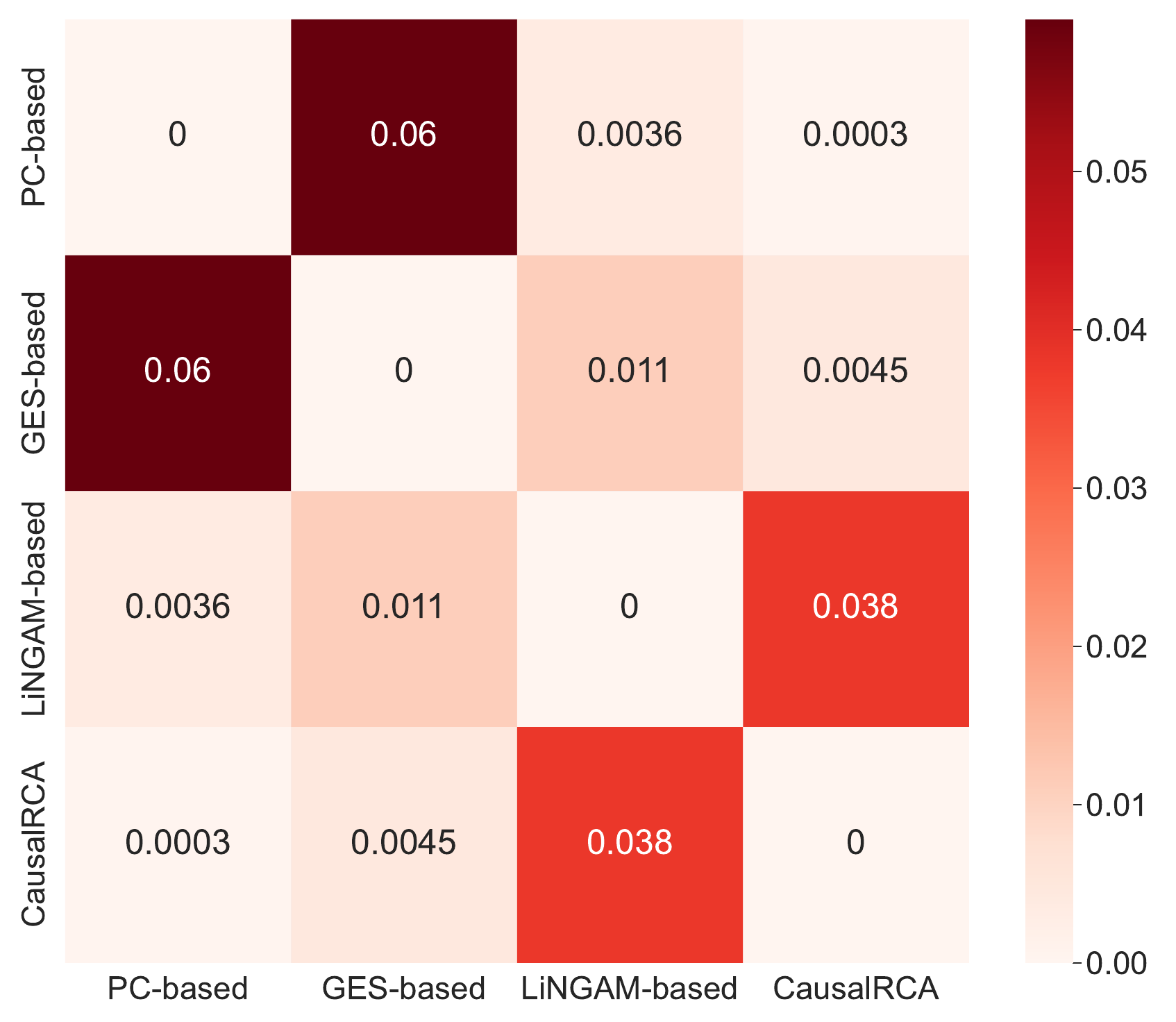}
\caption{\label{fig:lat_pvalue} P-value of RCA methods (Coarse-grained experiment)}
\end{minipage}
\begin{minipage}[t]{0.4\textwidth}
\centering
\includegraphics[width=7.3cm]{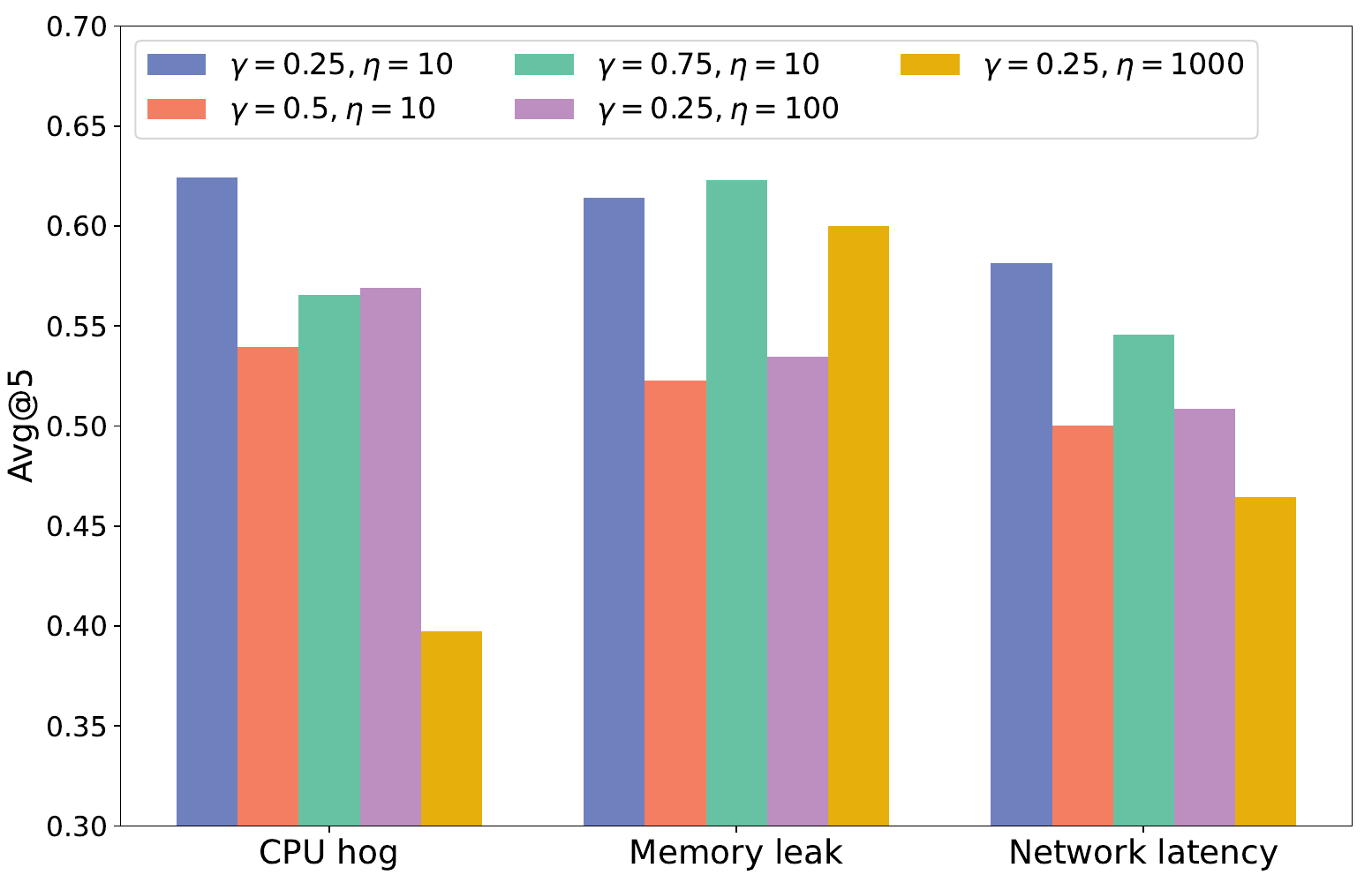}
\caption{\label{fig:lat_g_e} Localization accuracy with different $\gamma$ and $\eta$ (Coarse-grained experiment)}
\end{minipage}
\end{figure}

We evaluate the performance of CausalRCA on localizing the faulty service that initiates performance anomalies. This localization is conducted based on service-level data, which is the latency of all services. Table \ref{tab:faulty-service} shows the localization accuracy compared with baseline methods for different anomalies. We can see that, when compared to baseline methods, CausalRCA has improved localization accuracy in terms of $AC@1$, $AC@3$, and $Avg@5$ in different anomalies by up to 10\%. In addition, for CPU hog, causalRCA has the best performance in terms of $AC@1$, $AC@3$, and $Avg@5$. The $AC@3$ is 0.7175, which means that there is a 71.75\% chance of finding the root cause in the top three metrics on the ranked list, which is slightly higher than the LiNGAM-based method. For memory leak, CausalRCA continues to outperform in terms of $AC@1$, $AC@3$, and $Avg@5$. There is a 62.14\% possibility of localizing the root cause in the top 3 metrics. For network delay, $AC@3$ is not good enough, but $AC@1$ and $Avg@5$ are higher than baseline methods. In general, the average $AC@1$ of CausalRCA is 0.2, which means that there is an average 20\% possibility that the top 1 metric on the ranked list can be identified as the root cause. The averages $AC@3$ and $Avg@5$ of CausalRCA for the three anomalies are 0.5749 and 0.5815, respectively. The increase of average $Avg@5$ is 6.72\%, showing the improvement in localizing accuracy compared with baseline methods. We provide statistical testing to show the significant difference between these RCA methods. We obtained a p-value of 0.0003 using the ANOVA method first, showing a significant performance difference between the four RCA methods. We further utilized t-tests to compare the performance differences between each pair of methods, and the resulting p-values are shown in Figure \ref{fig:lat_pvalue}. We can see that CausalRCA has a significant difference from baseline methods, while PC-based and GES-based methods have no significant difference. 

\begin{table}[htb]
\centering
\caption{\label{tab:faulty-service} Localization accuracy of CI-based methods on localizing faulty services (Coarse-grained) for different anomalies} 
\scalebox{0.75}{
\begin{tabular}{|c|c|c|c|c|c|}
\hline
\multicolumn{1}{|c|}{\textbf{Methods}} &
\multicolumn{1}{c|}{\textbf{PC-based}} & 
\multicolumn{1}{c|}{\textbf{GES-based}} &
\multicolumn{1}{c|}{\textbf{LiNGAM-based}} &
\multicolumn{1}{c|}{\textbf{CausalRCA}} &
\multicolumn{1}{c|}{\textbf{Increase}} 
\\ \hline
\multicolumn{6}{|l|}{\textbf{CPU hog}} \\ \hline 
AC@1 & 0.1429 & 0.1429 & 0.1429 & \textbf{0.1873} & 4.44\% \\ \hline 
AC@3 & 0.2857 & 0.4286 & 0.7143 & \textbf{0.7175} & 0.32\%\\ \hline 
\textit{Avg@5} & 0.4286 & 0.4 & 0.5714 & \textbf{0.6244} & 5.3\%\\ \hline 
\multicolumn{6}{|l|}{\textbf{Memory leak}} \\ \hline 
AC@1 & 0 & 0 & 0.1429 & \textbf{0.2429} & 10\%\\ \hline 
AC@3 & 0.4286 & 0.1429 & 0.5714 & \textbf{0.6214} &5\% \\ \hline 
\emph{Avg@5} & 0.4286 & 0.2286 & 0.5429 & \textbf{0.6143} & 7.14\%\\ \hline 
\multicolumn{6}{|l|}{\textbf{Network delay}}  \\ \hline
AC@1 & 0.1429 & 0 & 0.1429 & \textbf{0.1714} & 2.85\% \\ \hline 
AC@3 & \textbf{0.5714} & 0 & 0.4286 & 0.3857 & -\\ \hline 
\emph{Avg@5} & 0.4857 & 0.1714 & 0.4286 & \textbf{0.5057} & 2\% \\ \hline 
\textbf{Average Avg@5} & 0.4476 & 0.2667 & 0.5143 & \textbf{0.5815} & 6.72\% \\ \hline 
\end{tabular}}
\end{table}

We analyze the impact of parameters $\gamma$ and $\eta$ in causal structure learning on the root cause localization performance of CausalRCA. We use $\gamma = 0.25$ and $eta = 10$ as default, and also set $\gamma = 0.5, 0.75$, and $\eta = 100, 1000$. The results can be found in Figure \ref{fig:lat_g_e}. We can see that $\gamma=0.25, \eta=10$ has the best performance in CPU hog and Network latency, and it also has the best average performance of different anomalies. In addition, $\gamma=0.75, \eta=10$ performs best for memory leak, because a high $\gamma$ can prevent too fast convergence and find better solutions, but it usually takes more time. Furthermore, we can see that $\gamma=0.25, \eta=1000$ performs poorly on CPU hog anomaly, but relatively well on memory leak anomaly, suggesting that a high $\eta$ can lead to more variance in localization results. 

\begin{figure}[!h]
\centering
\subfigure[CPU hog]{
    \includegraphics[width=0.22\textwidth]{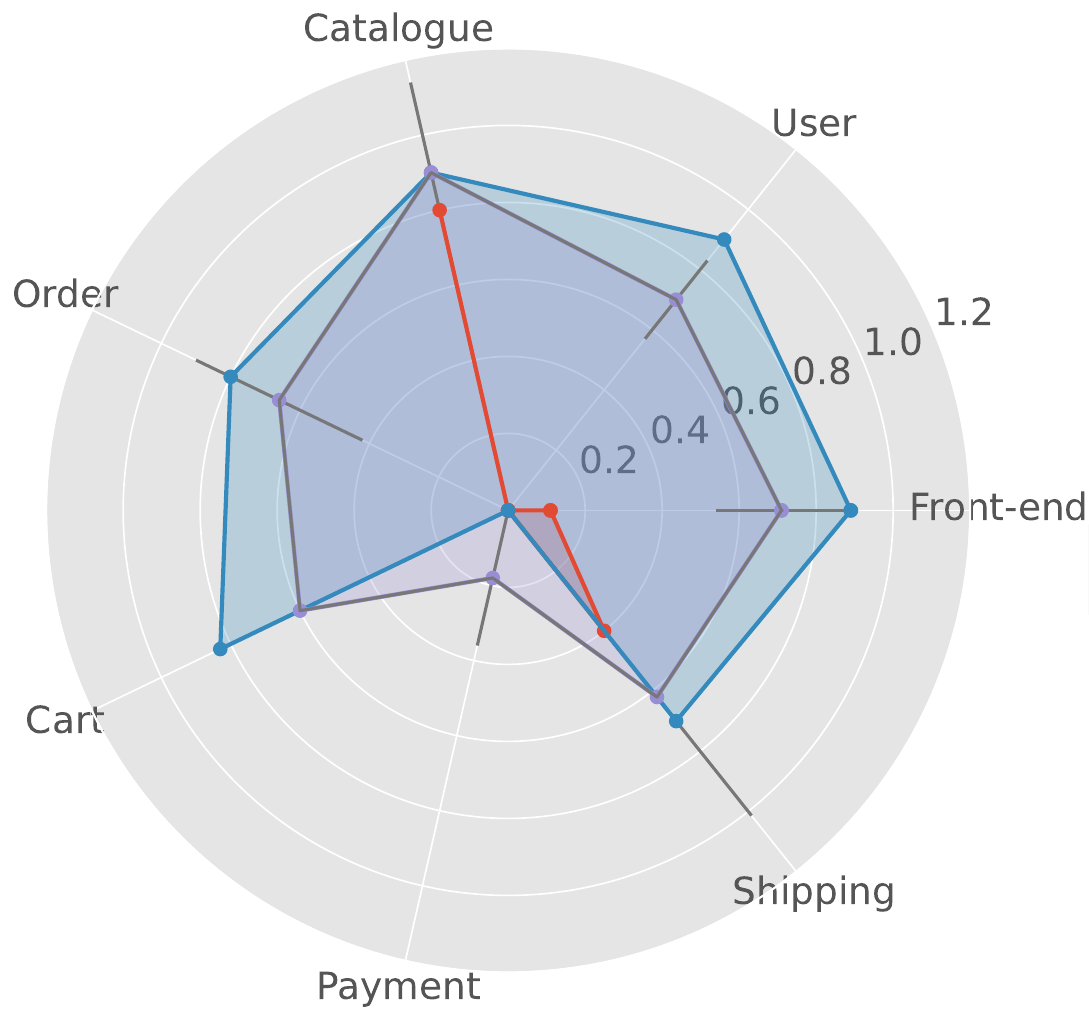}
}
\subfigure[Memory leak]{
\includegraphics[width=0.22\textwidth]{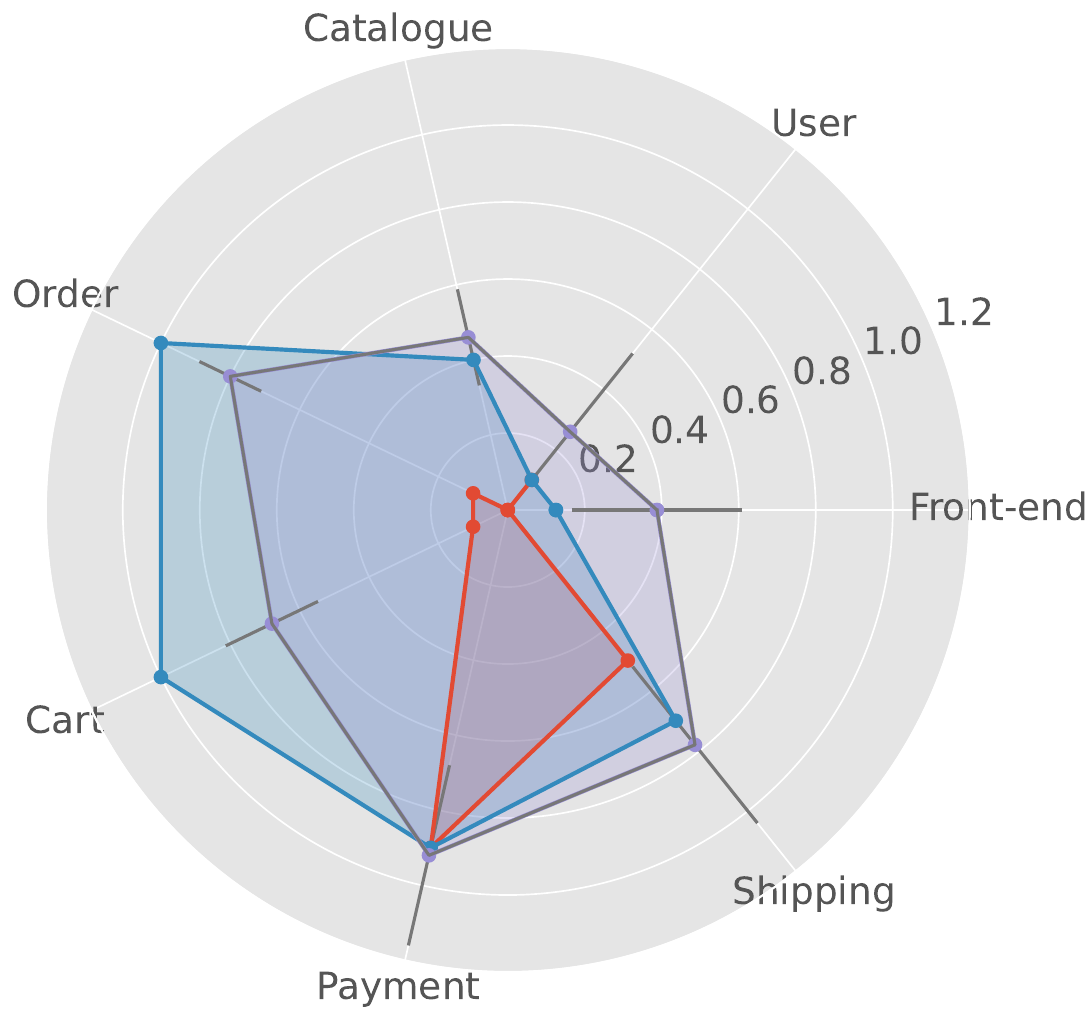}
}
\subfigure[Network delay]{
	\includegraphics[width=0.27\textwidth]{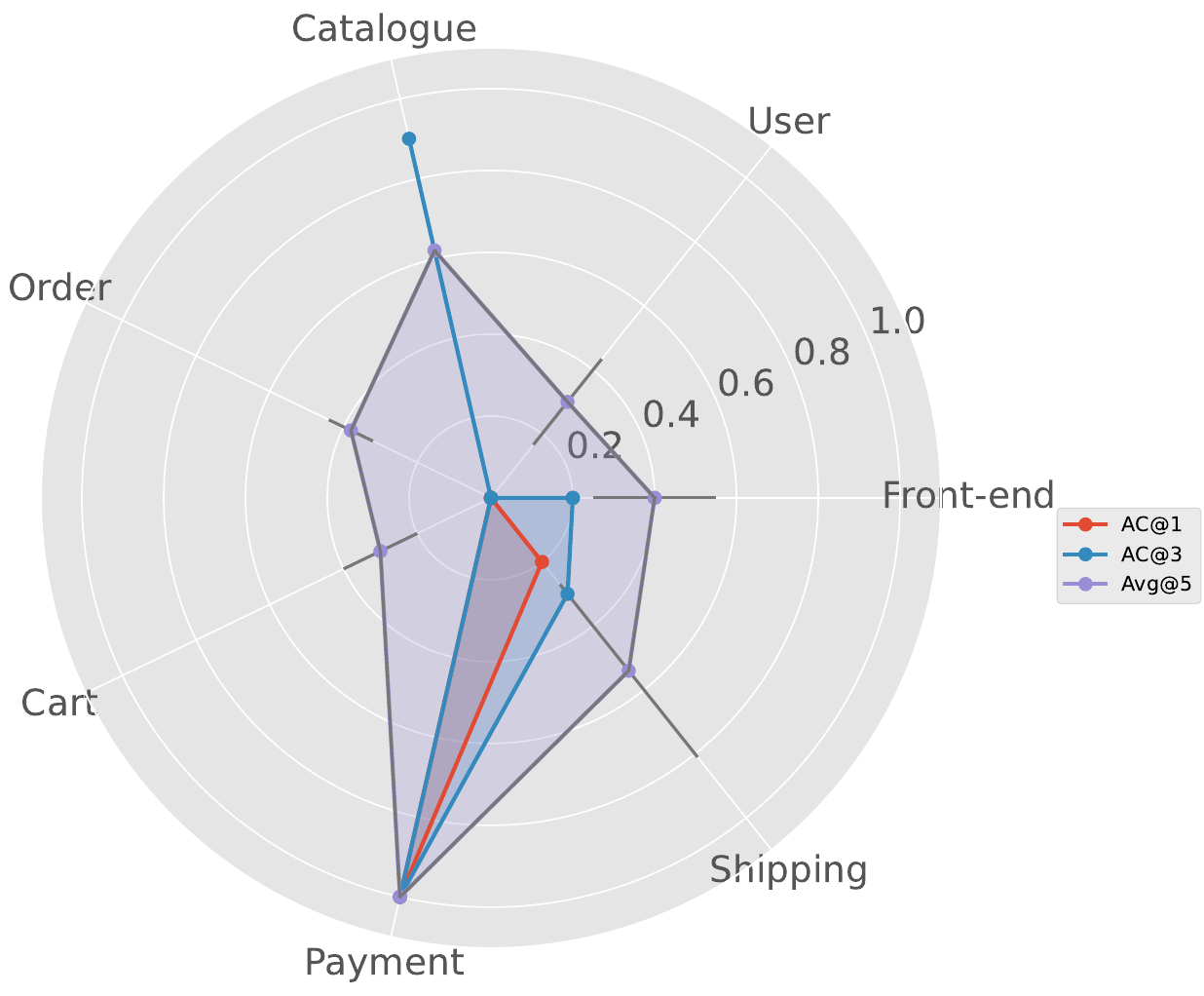}
}
\caption{Performance of CausalRCA on localizing faulty services with different anomalies}
\label{fig:serv-perf-lat}
\end{figure}

We then provide a detailed performance of CausalRCA on localizing faulty services with different anomalies in Figure \ref{fig:serv-perf-lat}. For CPU hog in Figure \ref{fig:serv-perf-lat}(a), we can see that localization accuracy performs well on services except $payment$, because $payment$ is not a CPU-intensive service. For the memory leak in Figure \ref{fig:serv-perf-lat}(b), we can see that $front$-$end$, $user$, and $catalogue$ perform worse than other services. The memory leak issues in these services do not affect their service latency much, making it difficult to identify cause-effect relations between services. In terms of network delay in Figure \ref{fig:serv-perf-lat}(c), only service $payment$ performs well, which explains the poor average localization performance of $AC@3$ in Table \ref{tab:faulty-service}. $Payment$ service relies heavily on the network, making it easy to localize the network delay issue. We plot the errorbar for $Avg@5$ in Figure \ref{fig:serv-perf-lat} to represent the variation of our results, and we can see that the standard deviations of many results are not high. 

\subsubsection{Fine-grained root cause metric localization in the faulty service}

Given the faulty service, we apply CausalRCA to container resource metrics and evaluate its performance on localization accuracy for different anomalies. Table \ref{tab:rca-metrics-false} shows the localization accuracy of CausalRCA on localizing root cause metric in faulty service compared with baseline methods. For different anomalies, we can see that CausalRCA has improved localization accuracy compared with baseline methods in terms of $AC@1$, $AC@3$, and $Avg@5$ by up to 14.29\%. In addition, for CPU hog, CausalRCA has the best performance in terms of $AC@3$ and $Avg@5$, while the $AC@1$ is worse than PC-based methods. For memory leak, the $AC@1$ of CausalRCA is worse than LiNGAM-based method, but it still has the best $AC@3$ and $Avg@5$. Finally, for network delay, CausalRCA outperforms in terms of $AC@1$, $AC@3$, and $Avg@5$. In general, the average $AC@1$ of CausalRCA is 0.2476, which means there is about a 25\% possibility of determining the top 1 metric on the ranked list as the root cause. For the three anomalies, the average $AC@3$ is 0.719, which means there is a 71.9\% possibility to localize the root cause metric in the top 3 metrics on the ranked list, and the average improvement is 10\% compared with baseline methods. The average $Avg@5$ of CausalRCA is 0.6681, and the average increase is 9.43\%. We consider the outperformance of CausalRCA is because resource metrics, such as CPU/memory usage, affect each other, which makes it easier to identify anomaly propagation with CI methods. 

\begin{table}[!h]
\centering
\caption{\label{tab:rca-metrics-false} Localization accuracy of CI-based methods on localizing root cause metrics (Fine-grained) in faulty services for different anomalies} 
\scalebox{0.75}{
\begin{tabular}{|c|c|c|c|c|c|}
\hline
\multicolumn{1}{|c|}{\textbf{Methods}} &
\multicolumn{1}{c|}{\textbf{PC-based}} & 
\multicolumn{1}{c|}{\textbf{GES-based}} &
\multicolumn{1}{c|}{\textbf{LiNGAM-based}} &
\multicolumn{1}{c|}{\textbf{CausalRCA}} &
\multicolumn{1}{c|}{\textbf{Increase}} 
\\ \hline
\multicolumn{6}{|l|}{\textbf{CPU hog}} \\ \hline 
AC@1 & \textbf{0.4286} & 0.1429 & 0 & 0.2286 & - \\ \hline 
AC@3 & 0.4286 & 0.5714 & 0.7143 & \textbf{0.7286} & 1.43\% \\ \hline 
\emph{Avg@5} & 0.4286 & 0.6 & 0.5714 & \textbf{0.67} & 7\% \\ \hline 
\multicolumn{6}{|l|}{\textbf{Memory leak}} \\ \hline 
AC@1 & 0 & 0 & \textbf{0.4286} & 0.2714 & - \\ \hline 
AC@3 & 0.1429 & 0.4286 & 0.5714 & \textbf{0.7143} & 14.29\% \\ \hline 
\emph{Avg@5} & 0.3429 & 0.4 & 0.6286 & \textbf{0.6771} & 4.85\% \\ \hline 
\multicolumn{6}{|l|}{\textbf{Network delay}} \\ \hline
AC@1 & 0 & 0.1429 & 0.1429 & \textbf{0.2429} &10\% \\ \hline 
AC@3 & 0.2857 & 0.5714 & 0.4286 & \textbf{0.7143} & 14.29\%\\ \hline 
\emph{Avg@5} & 0.2214 & 0.5143 & 0.5214 & \textbf{0.6571} & 13.57\% \\ \hline 
\textbf{Average Avg@5} & 0.33 & 0.5048 & 0.5738 & \textbf{0.6681} & 9.43\% \\ \hline 

\end{tabular}}
\end{table}

\begin{figure}[htbp]
\centering
\begin{minipage}[t]{0.4\textwidth}
\centering
\includegraphics[width=6cm]{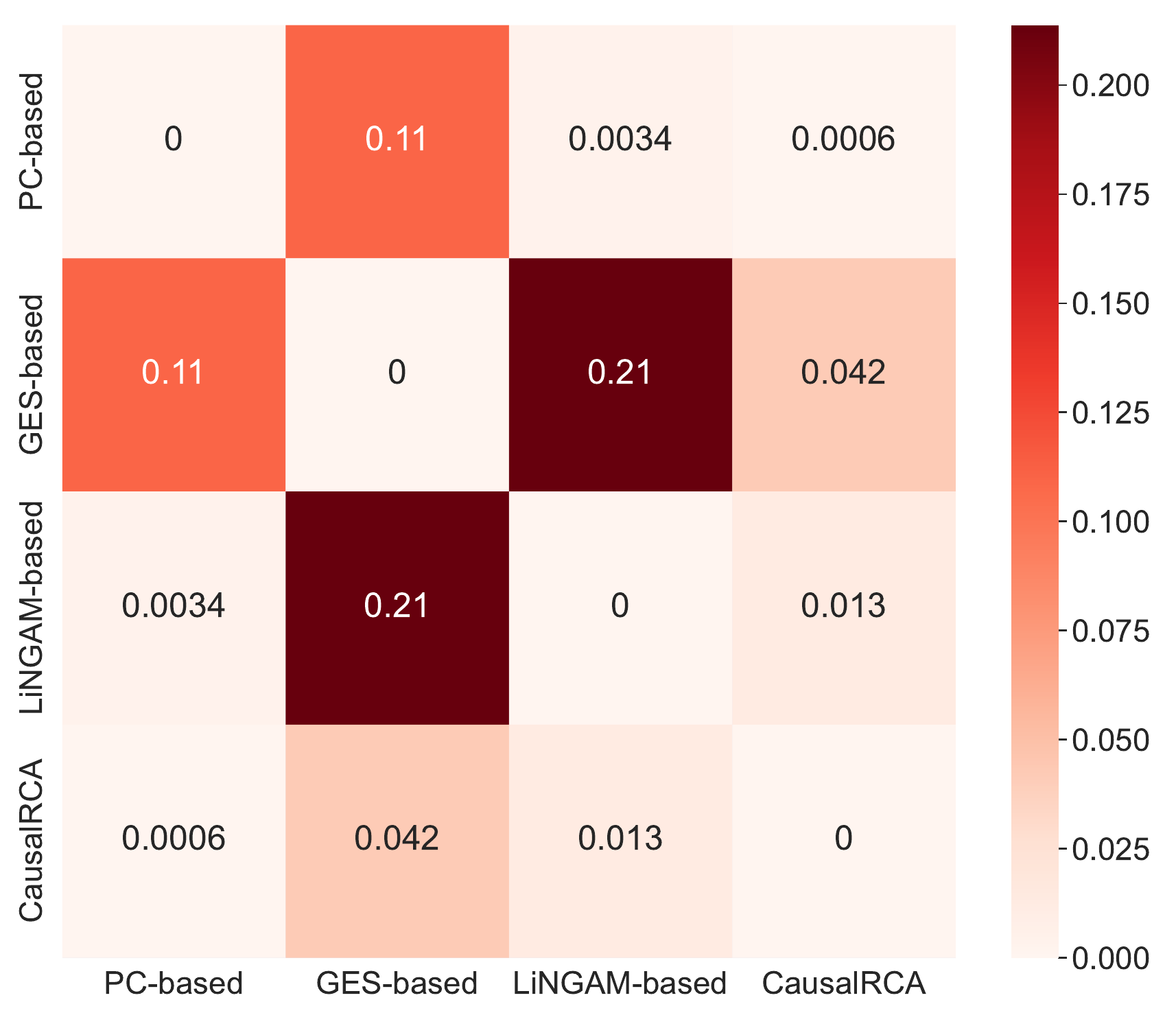}
\caption{\label{fig:sig_pvalue} P-value of RCA methods (Fine-grained experiment)}
\end{minipage}
\begin{minipage}[t]{0.4\textwidth}
\centering
\includegraphics[width=7.3cm]{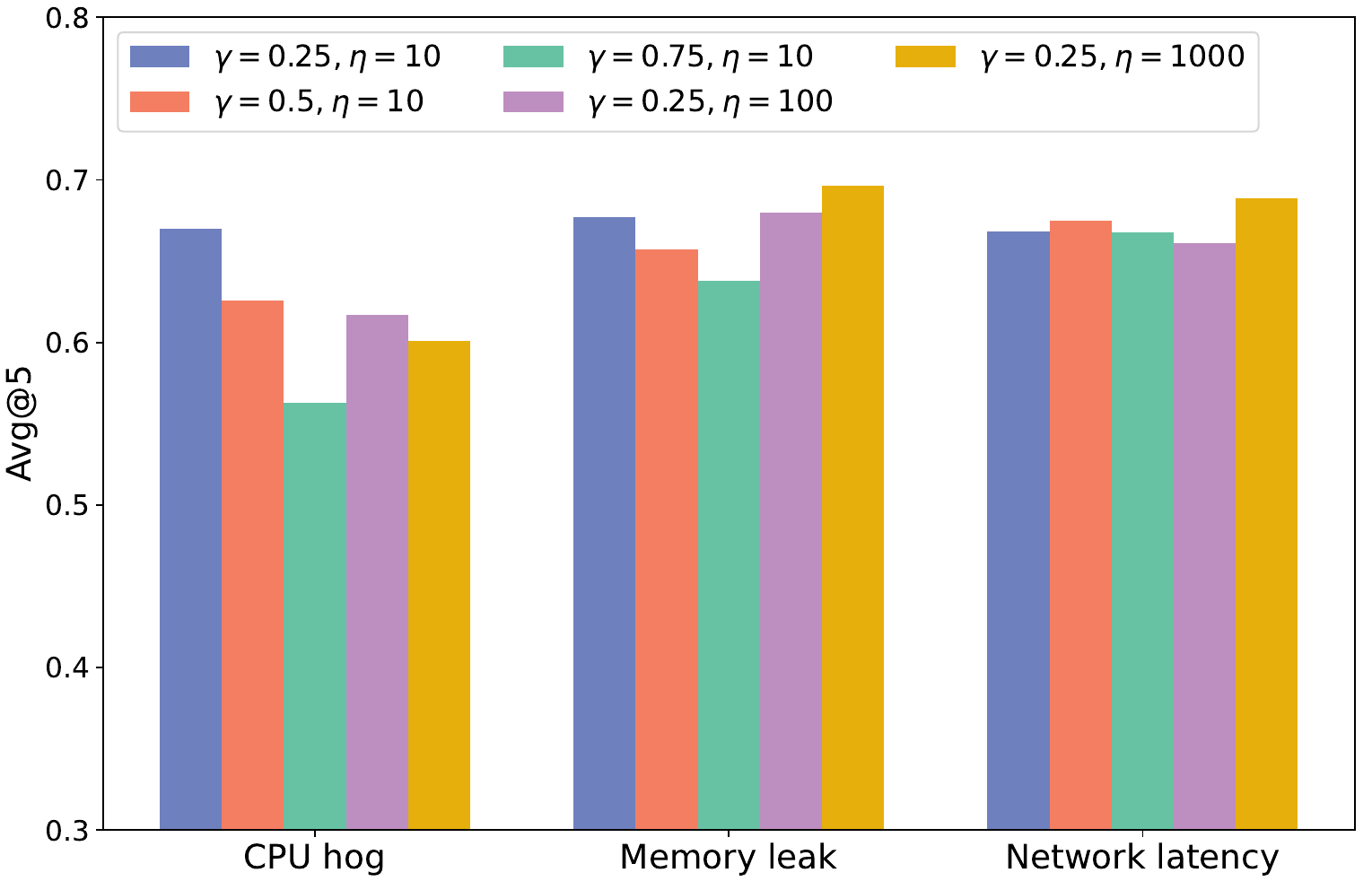}
\caption{\label{fig:sig_g_e} Localization accuracy with different $\gamma$ and $\eta$ (Fine-grained experiment)}
\end{minipage}
\end{figure}

We also provide statistical testing to show the significant difference of these RCA methods. We first obtained a p-value of 0.0013 using the ANOVA method, showing a significant performance difference between the four RCA methods. The p-values obtained from t-tests are shown in Figure \ref{fig:sig_pvalue}. We can see that CausalRCA has a significant difference from baseline methods. In comparison, GES-based method has no significant difference with PC-based and LiNGAM-based methods.  

We evaluate the impact of parameters $\gamma$ and $\eta$ and present the findings in Figure \ref{fig:sig_g_e}. The results indicate that $\gamma=0.25$ and $\eta=10$ perform the best in identifying CPU hog anomalies, while $\gamma=0.25$ and $\eta=100$ are most effective in detecting memory leak and network latency anomalies. In addition, $\gamma=0.5$ and $\eta=10$ outperform $\gamma=0.5$ and $\eta=100$ in detecting network latency anomalies and perform better than $\gamma=0.75$ and $\eta=10$ across all three types of anomalies. Overall, $\gamma=0.25$ and $\eta=10$ have the highest average localization performance. However, increasing $\gamma$ and $\eta$ could potentially lead to better solutions and improve localization accuracy.

\begin{figure}[!h]
\centering
\subfigure[AC@3]{
    \includegraphics[width=0.4\textwidth]{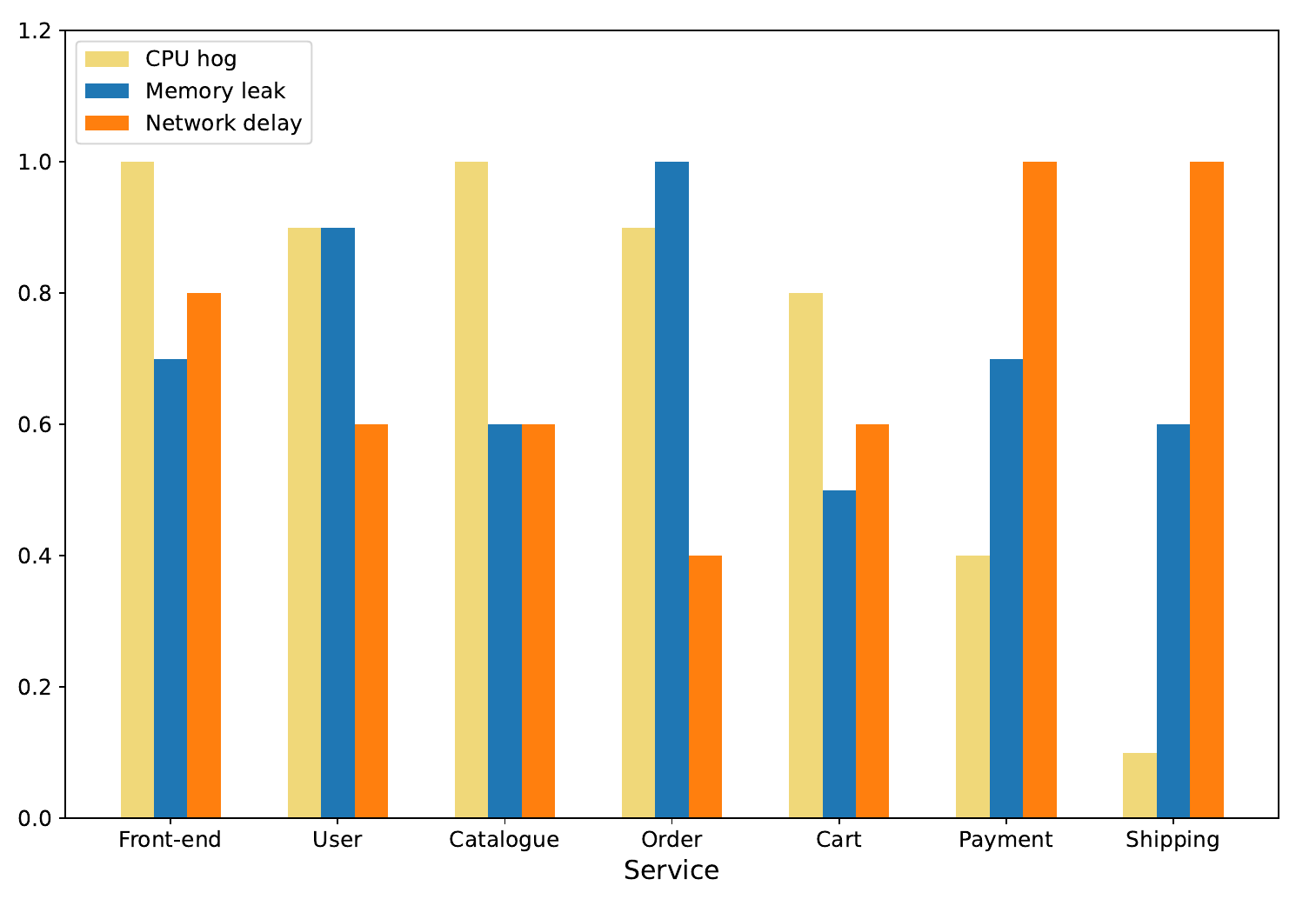}
}
\subfigure[Avg@5]{
\includegraphics[width=0.4\textwidth]{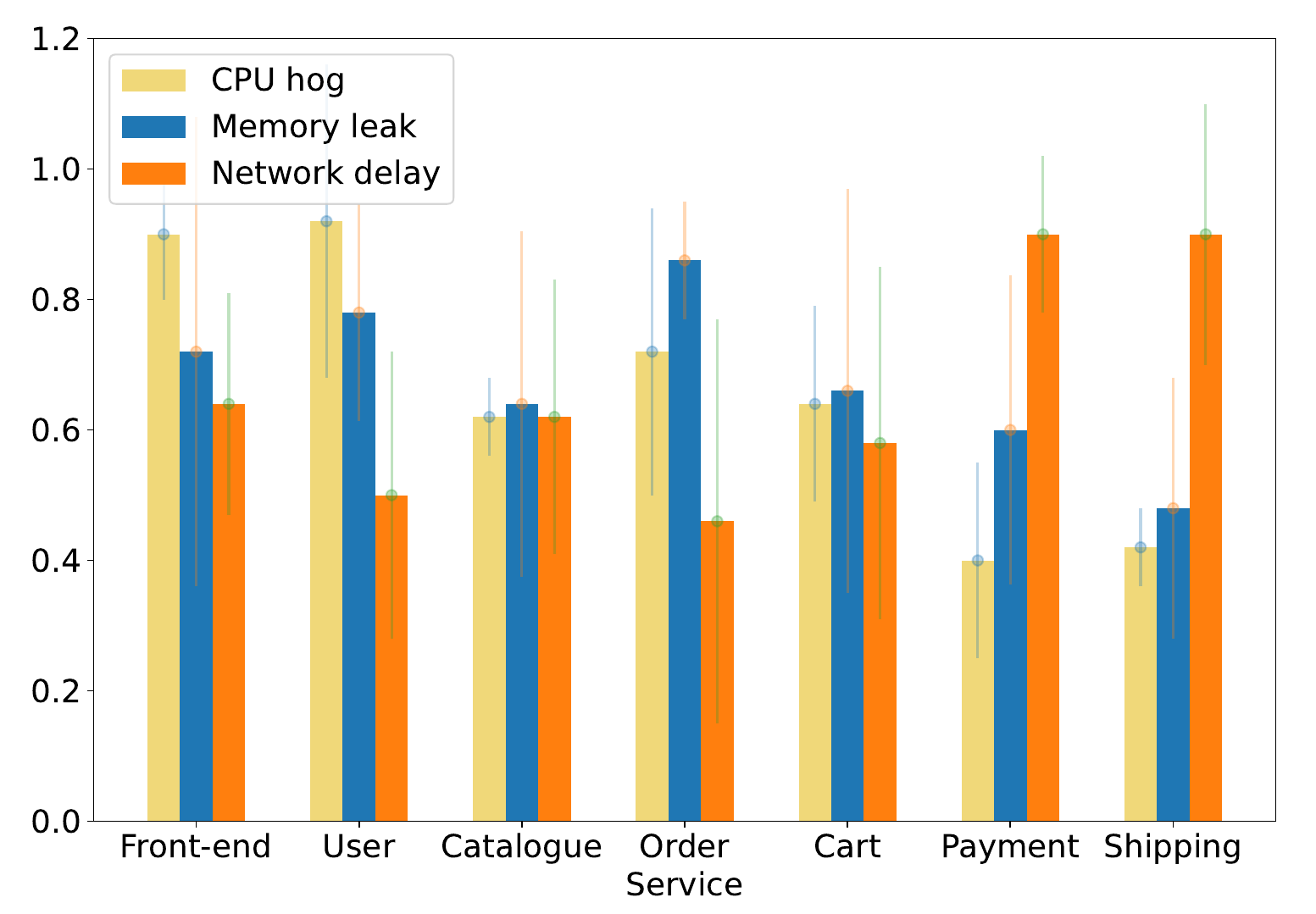}
}
\caption{Performance of CausalRCA on localizing root cause metrics in faulty services with different anomalies}
\label{fig:ac35-false}
\end{figure}

We then provide the performance of CausalRCA on localizing root cause metrics in faulty services with the three anomalies in Figure \ref{fig:ac35-false}. We can see that $Ac@3$ and $AC@5$ have consistent performance. For CPU hog, we can see that localization accuracy is low for $payment$ and $shipping$ services because they are insensitive to CPU resources. Because the memory leak issue manifests in multiple resource metrics, all services perform well for the memory leak. $Order$ service has the best performance because it is highly related to memory usage. For network delay, we can see that $payment$ and $shipping$ have the best performance because they rely heavily on the network. We plot the errorbar for $Avg@5$ in Figure \ref{fig:ac35-false}(b), and we can see there are some variances in the localization results, maybe caused by the dynamic nature of cloud environments or random fluctuation of resources in services, showing that the generality of CausalRCA can be explored more in the future. 

\subsubsection{Fine-grained root cause metric localization with all monitoring metrics}

\

\begin{figure}[!h]
\centering
\subfigure[CPU hog]{
    \includegraphics[width=0.37\textwidth]{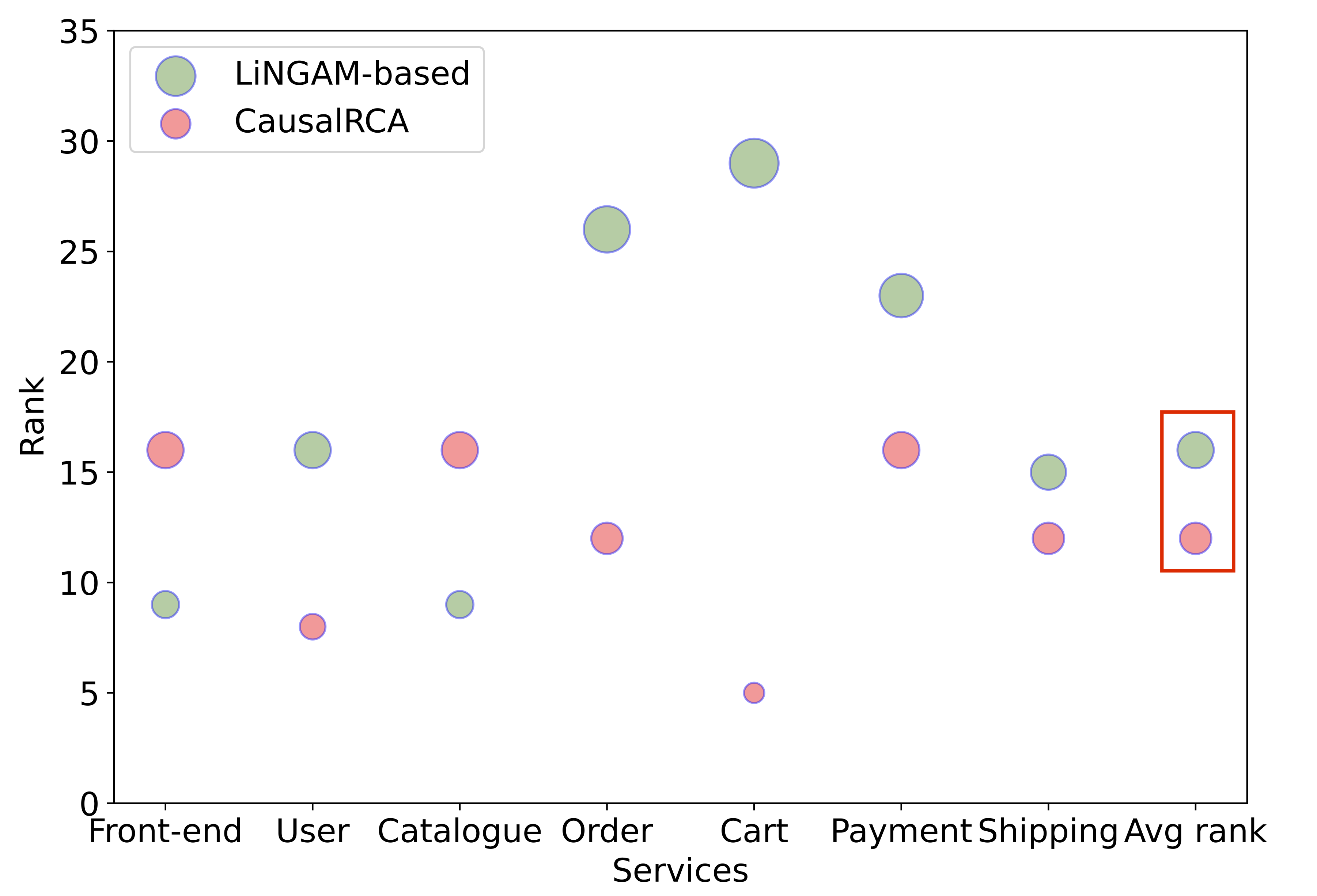}
}
\subfigure[Memory leak]{
\includegraphics[width=0.37\textwidth]{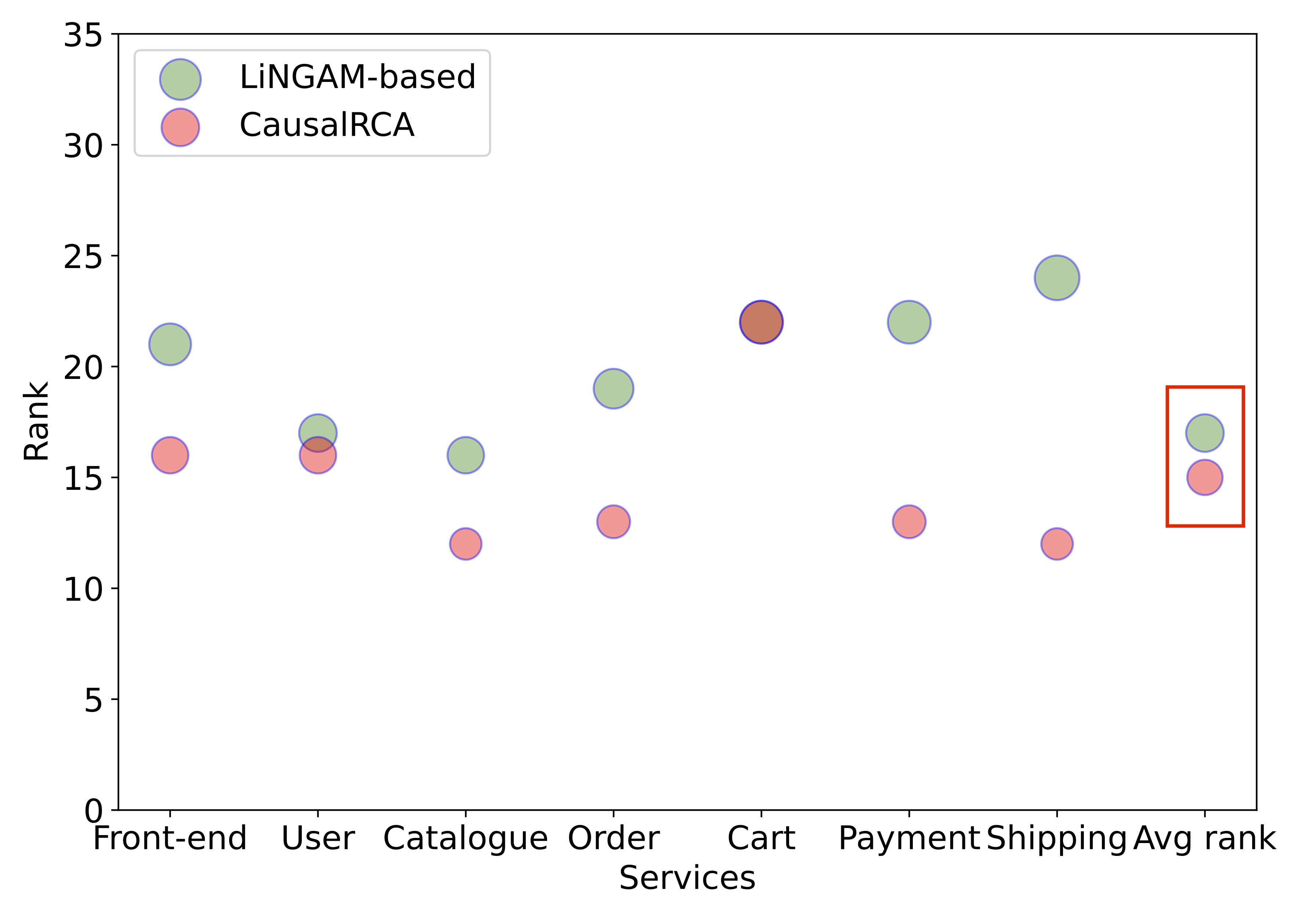}
}
\subfigure[Network delay]{
	\includegraphics[width=0.37\textwidth]{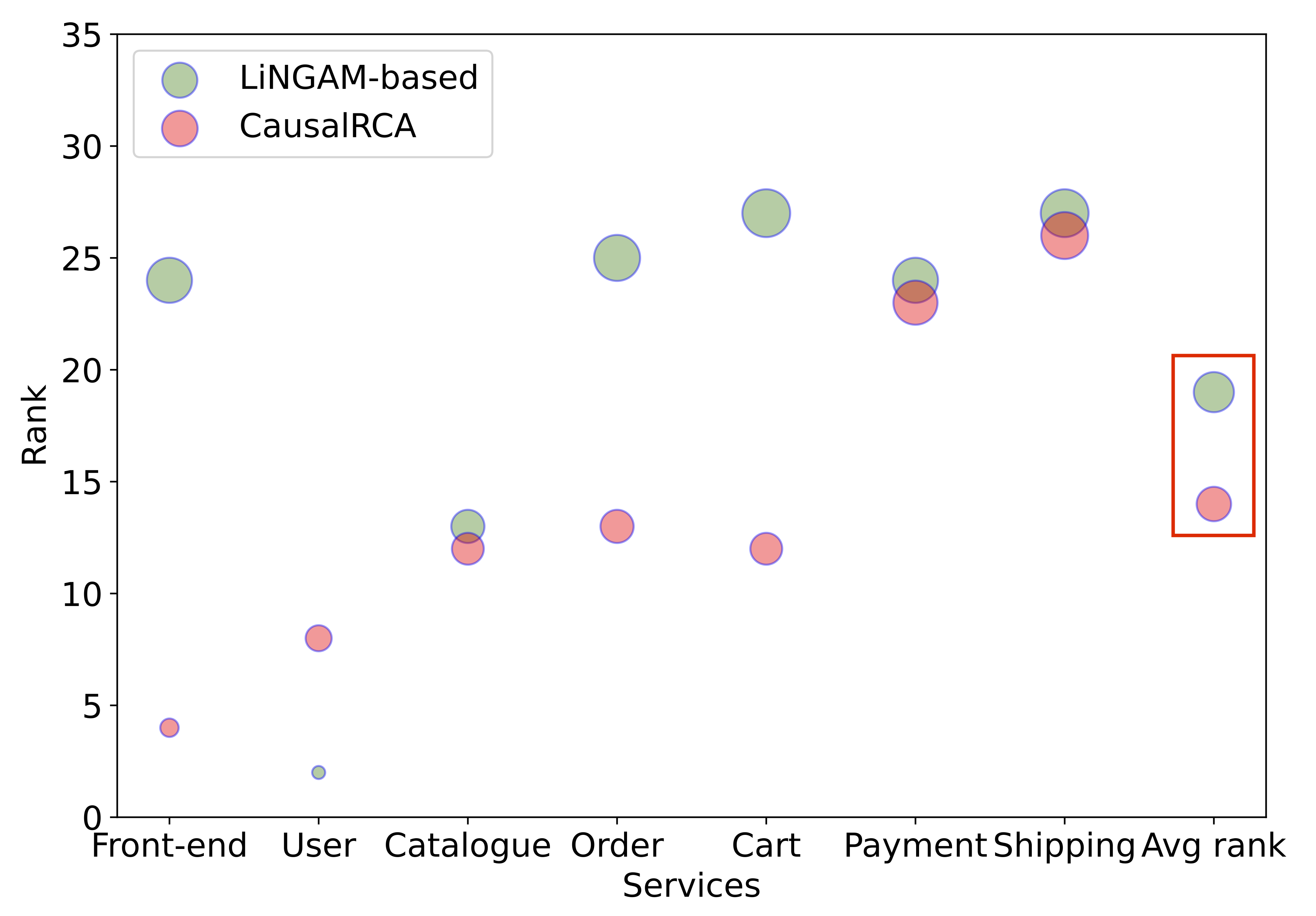}
}
\caption{Ranks of root cause metrics identified by CI-based methods}
\label{fig:all-service}
\end{figure}

\begin{figure}[htbp]
\centering
\includegraphics[width=0.4\textwidth]{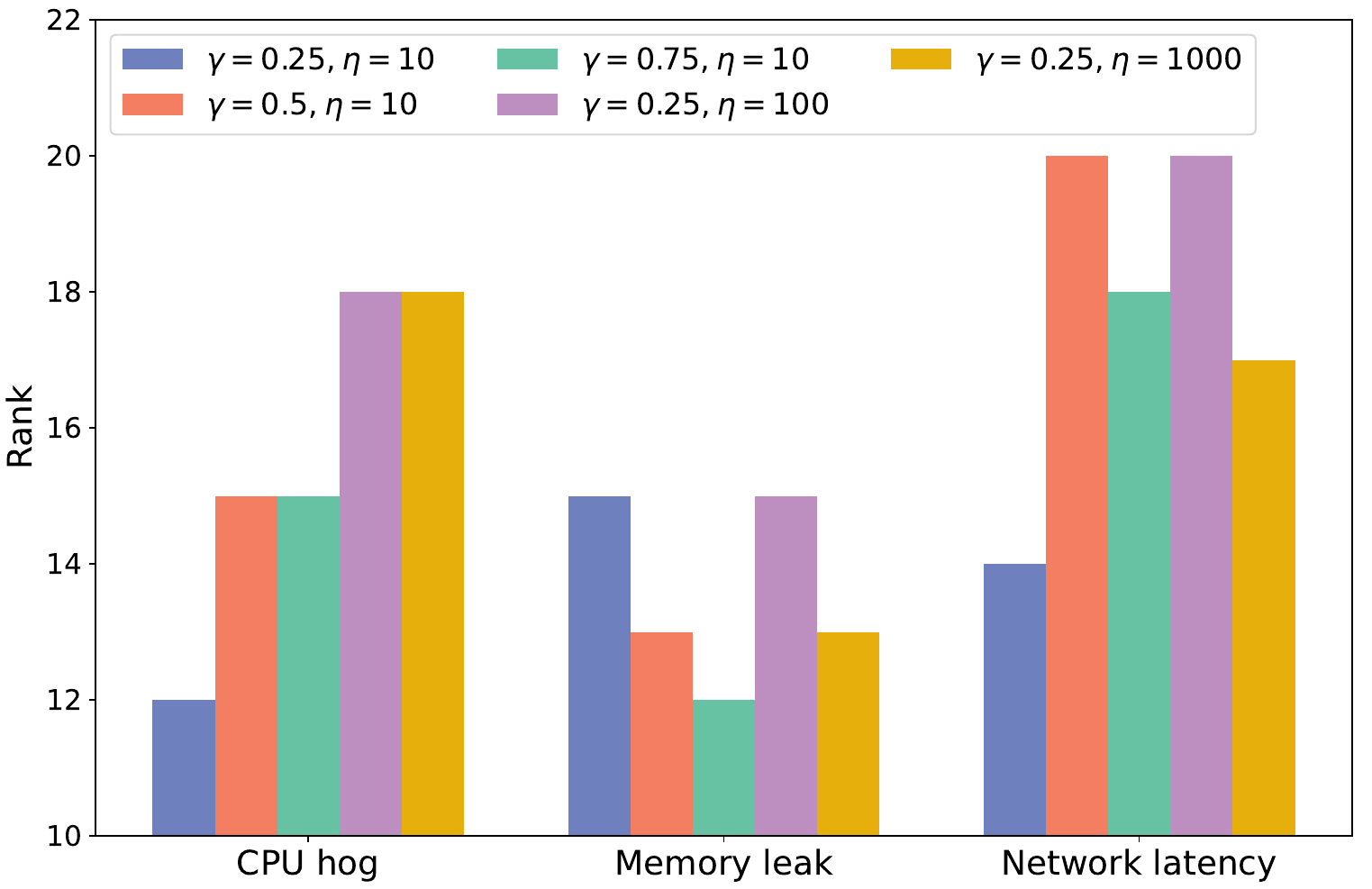}
\caption{Ranks of root cause metrics with different parameters $\gamma$ and $\eta$}
\label{fig:alls_g_e}
\end{figure}

%B. Combine latency graph with service graphs
%C. Combine arc graph with service graphs
Considering that we are unknown to services and underlying infrastructures of an application, we conduct the fine-grained root cause localization with all monitoring metrics. We apply CausalRCA on all monitoring metrics to localize the root cause metric. We mainly show the ranks of comparison between the LiNGAM-based method and CausalRCA, because PC sometimes fails to extract causal relations between metrics, while GES takes too long to build a causal graph with too many nodes. For this fine-grained root cause localization, it is hard to identify the root cause metric in the top 1 or top 3 metrics, so we use the rank of root cause metrics to evaluate localization performance as shown in Figure \ref{fig:all-service}. We can see that the average rank of CausalRCA is about 13, which is lower than the LiNGAM-based method. The result shows that CausalRCA has better localization performance than the LiNGAM-based method. However, it is still challenging to extract causal relations between metrics and pinpoint the root cause metric from multiple observable metrics. We apply the t-test to LiNGAM-based and CausalRCA methods and obtain the p-value of 0.0335, showing the significant difference between them. The impact of parameters $\gamma$ and $\eta$ is presented in Figure \ref{fig:alls_g_e}. The results indicate that $\gamma=0.25, \eta=10$ performs the best in identifying CPU hog and network latency anomalies. For the memory leak anomaly, higher $\gamma$ and $\eta$ have better localization accuracy, and $\gamma=0.75, \eta=10$ is most effective. On average, the $\gamma=0.25, \eta=10$ has the best localization performance, while adjusting $\gamma$ and $\eta$ for the memory leak anomaly can improve localization accuracy effectively. 

In conclusion, CausalRCA has a significant difference from baseline methods and better localization accuracy for coarse-grained and fine-grained root cause localization. In addition, we find that $\gamma=0.25, \eta=10$ in CausalRCA has the best localization performance on average. However, adjusting parameters can provide more potential for improving localization accuracy. Based on CI methods, we can see that anomaly propagation performs differently in different services; for example, the $payment$ service is sensitive to the network delay issue but nonsensitive to the CPU hog issue. As for fine-grained root cause localization, it is still challenging to pinpoint the root cause metric in all monitoring metrics. Therefore, it is more practical to consider the drill-down localization, i.e., identify the faulty service first and then determine the root cause metric in the faulty service. 

\subsection{Threats to validity}
We analyze threats to our framework from the four categories: construct, internal, conclusion, and external validity based on \cite{wohlin2012experimentation}. The \textbf{construct threat} to validity mainly lies in the hyperparameters and evaluation metrics. We provide parameter analysis for two hyperparameters in CausalRCA, and results show that default parameters works well as provided in \cite{yu2019dag} but tuning parameters carefully has the potential of improving localization accuracy. In addition, we use widely used evaluation metrics and provide statistical testing to evaluate the performance difference of different RCA methods. 

The \textbf{internal threat} to validity mainly lies in the implementation of the framework, as errors or bugs in the implementation could affect the accuracy of the results. To reduce it, we have used established Python packages and conducted thorough testing. We have also repeated the experiments multiple times to ensure the reliability and consistency of our results. The \textbf{conclusion threat} to validity of our framework is related to the types of anomalies used in experiments. As microservice applications have a variety of performance anomalies that can affect the localization results \cite{mariani2018localizing}, we injected three different types of common and frequent anomalies to evaluate the effectiveness of our framework. We report and discuss the localization results for each individual anomaly type, and the experimental results demonstrate the superior performance of our framework on these anomalies. 

The \textbf{external threat} relies on the configuration of microservice applications and the data collection strategies. In this paper, we investigate a specific configuration of a microservice application to evaluate the performance of CausalRCA, which may limit the generality of our framework. However, building complex infrastructures and repeating the experiments on multiple testbeds is extremely expensive, which is impractical for our experiments. In addition, the benchmark microservice application sock-shop is widely used in academia to aid the testing of microservices in clouds \cite{lin2018microscope, wu2020microrca}, and it helps us mitigate this threat. On the other hand, the localization performance of our framework heavily relies on input data. To mitigate the threat, we adopt Prometheus, an open-source tool for real-time monitoring, and collect service-level and resource-level metrics that present the status of a running microservice application. Currently, our framework performs well with anomalies injected over fixed time range anomalies, but the effect of different time ranges for CausalRCA can be explored more.

\section{Discussion}
\label{dis}
This paper provides a framework called CausalRCA for root cause localization of microservice applications. The framework is developed with CI-based methods, including causal structure learning and root cause inference. We provide coarse-grained and fine-grained experiments to evaluate the localization performance of the framework. Our experimental results show that the framework has the best localization accuracy compared with baseline methods. However, some aspects can still be improved.

Our experiments show that CausalRCA performs well on localizing faulty services and root cause metrics in faulty services. However, localizing root cause metrics from all monitoring metrics is very hard. The average rank of root cause metrics is out of ten. We consider the improvement of localization accuracy can be researched more. First, data preprocessing, such as feature reduction, can be considered to reduce training time and improve localization accuracy. Next, we apply a gradient-based method to learn causal structures. The gradient-based method is applied to time-series monitoring data, which may ignore time lags in the original data. We consider that time lags in the data may help improve causal structure learning. For the root cause inference method PageRank, a personalized PageRank \cite{jeh2003scaling}, which considers the preferences of nodes, can be applied. 

This paper mainly focuses on monitoring data to implement root cause localization. Monitoring data has multi-dimensional information and is easy to collect compared with trace and log data. However, trace data and log data contain accurate deployment information and service interactions, which can be used to calibrate the causal graph generated based on monitoring data. At the same time, the causal graph generated with CI methods can extract hidden relations between metrics. Therefore, we can consider combining different data resources to improve localization accuracy in the future.

This paper mainly focuses on improving localization accuracy of microservices, while efficiency is also important for achieving fast recovery. For the sock-shop benchmark application in this paper, we roughly estimate the time spent of our framework takes tens seconds, showing the cost of time may be lower than service migration \cite{chen2022dynamic}. However, for large-scale microservice, e.g., hundreds/thousands of services, the time cost for building the causality graph may be far greater than service migration. We will test the scalability and exact time cost of our framework and pay more attention to reducing training time in the future. For now, we mainly use the data collected in five minutes after the anomaly is detected. In the future, we will consider testing localization performance with different time ranges based on our CausalRCA.

\section{Conclusion}
\label{con}
This paper tackles the challenge of localizing root causes of performance anomalies in microservice applications. Root cause localization can be used to help operators achieve fast recovery of microservice applications. Therefore, it is important to guarantee localization accuracy at first. In addition, fine-grained root cause localization, which means identifying both the faulty service and resource related metrics in a faulty service, is necessary. With monitoring data, we provide a CI-based framework named CausalRCA, which can automate localizing root causes with fine granularity and in real time. 

The CausalRCA works with causal structure learning and root cause inference components. For causal structure learning, we propose a GNN-based method that uses a deep generative model and applies a variant of the structural constraint to learn the weighted DAG. Compared with other CI methods, the gradient-based method can extract non-linear causal relations between metrics. For root cause inference, we apply PageRank to visit the weighted DAG and return a ranked list of all metrics. We then provide experiments to evaluate the localization performance of CausalRCA.  

In our experiments, we conduct three types of experiments: coarse-grained faulty service localization, fine-grained root cause metrics localization in faulty services, and fine-grained root cause metrics localization with all monitoring metrics. Our experimental results show that CausalRCA has better localization accuracy compared with baseline methods. Furthermore, based on CI methods, we can see that anomaly propagation performs differently between services, which gives operators a better understanding of microservice applications. In addition, it is difficult for fine-grained root cause localization with all monitoring metrics because anomalous metrics manifest diverse symptoms in different services. Therefore, fine-grained root cause localization with all monitoring metrics still needs to be improved. However, for microservice applications, we can still consider the drill-down localization, first identifying the faulty service, then pinpointing the root cause metric in the faulty service to identify fine-grained causes. 

In the future, we will continue to improve the localization performance of CausalRCA. Hyperparameters tuning can be tested more in the future. The causal structure learning can consider time lag in monitoring data, and the root cause inference can be improved by adding the preferences of nodes. In addition, employing knowledge from trace and log data to calibrate the causal graph may improve localization accuracy and make the causal graph more reasonable. Finally, localization efficiency to achieve fast recovery needs to be tested and improved. 

\section{Acknowledge}
This research work is funded by the EU Horizon 2020 and Horizon Europe research and innovation program under grant agreements 825134 (ARTICONF project), 862409 (BlueCloud project), 824068 (ENVRIFAIR project) and 101094227 (BlueCloud 2026). The work is also supported by LifeWatch ERIC and partially funded by the Science and Technology Program of Sichuan Province under Grant 2020JDRC0067 and 2020YFG0326. 

\bibliographystyle{elsarticle-num} 
\bibliography{draft}

\end{document}